\documentclass[apj]{emulateapj}

\def\ks{km~s$^{-1}$}
\def\ms{m~s$^{-1}$}

\def\mjup{M$_{\rm Jup}$}
\def\mearth{M$_{\earth}$}
\def\msun{M$_{\odot}$}
\def\rsun{R$_{\odot}$}
\def\lsun{R$_{\odot}$}
\def\msini{$M_P\sin i~$}

\def\chis{$\chi^2_\nu$}

\def\feh{[Fe/H]}

\def\shk{$S_{\rm HK}$}
\def\logg{$\log{g}$}
\def\teff{T$_{\rm eff}$}
\def\vsini{$V_r\sin{i}$}
\def\caii{Ca\,II}

\def\starB{HD\,5891}
\def\pB{177.11}
\def\peB{0.31}

\def\tpB{15432}
\def\tpeB{10}
\def\eB{0.066}
\def\eeB{0.020}

\def\omB{360}
\def\omeB{30}
\def\kB{178.5}
\def\keB{4.1}
\def\jittB{17.6}
\def\jitteB{0.8}
\def\msiniB{7.6}
\def\msinieB{0.4}
\def\arelB{0.76}
\def\areleB{0.02}
\def\rmsB{28.4}

\def\nobsB{54}
\def\mstarB{1.91}
\def\mstareB{0.13}
\def\bvB{0.99}

\def\vmagB{8.25}

\def\mvB{1.3}
\def\mveB{0.7}
\def\vsiniB{4.95}
\def\ageB{1.5}
\def\ageeB{0.8}
\def\rstarB{8.7}
\def\rstareB{0.2}
\def\lstarB{39.4}
\def\lstareB{0.8}
\def\teffB{4907}
\def\loggB{2.9}
\def\feB{-0.02}

\def\dB{251}
\def\deB{76}
\def\starC{HD\,18742}
\def\pC{772}
\def\peC{11}

\def\tpC{15200}
\def\tpeC{110}

\def\eupC{0.23}
\def\omC{102}
\def\omeC{50}
\def\kC{44.3}
\def\keC{3.8}
\def\trendC{4.1}
\def\trendeC{1.6}

\def\jittC{7.6}
\def\jitteC{0.9}
\def\msiniC{2.7}
\def\msinieC{0.3}
\def\arelC{1.92}
\def\areleC{0.05}
\def\rmsC{7.9}

\def\nobsC{26}
\def\mstarC{1.60}
\def\mstareC{0.11}
\def\bvC{0.94}

\def\vmagC{7.97}

\def\mvC{2.3}
\def\mveC{0.2}
\def\vsiniC{2.98}
\def\ageC{2.3}
\def\ageeC{0.5}
\def\rstarC{4.9}
\def\rstareC{0.1}
\def\lstarC{13.9}
\def\lstareC{0.5}
\def\teffC{5048}
\def\loggC{3.3}
\def\feC{-0.04}

\def\dC{135}
\def\deC{14}
\def\starD{HD\,28678}
\def\pD{387.1}
\def\peD{4.2}

\def\tpD{15517}
\def\tpeD{30}
\def\eD{0.168}
\def\eeD{0.068}

\def\omD{131}
\def\omeD{30}
\def\kD{33.5}
\def\keD{2.3}
\def\trendD{3.1}
\def\trendeD{1.9}

\def\jittD{6.1}
\def\jitteD{0.8}
\def\msiniD{1.7}
\def\msinieD{0.1}
\def\arelD{1.24}
\def\areleD{0.03}
\def\rmsD{6.1}

\def\nobsD{30}
\def\mstarD{1.74}
\def\mstareD{0.12}
\def\bvD{1.01}

\def\vmagD{8.54}

\def\mvD{1.8}
\def\mveD{0.5}
\def\vsiniD{2.97}
\def\ageD{1.8}
\def\ageeD{0.7}
\def\rstarD{6.2}
\def\rstareD{0.1}
\def\lstarD{22.9}
\def\lstareD{0.6}
\def\teffD{5076}
\def\loggD{3.3}
\def\feD{-0.11}

\def\dD{227}
\def\deD{48}
\def\starE{HD\,33142}
\def\pE{326.6}
\def\peE{3.9}

\def\tpE{15324}
\def\tpeE{60}

\def\eupE{0.22}
\def\omE{138}
\def\omeE{60}
\def\kE{30.4}
\def\keE{2.5}
\def\jittE{7.6}
\def\jitteE{0.8}
\def\msiniE{1.3}
\def\msinieE{0.1}
\def\arelE{1.06}
\def\areleE{0.03}
\def\rmsE{8.3}

\def\nobsE{33}
\def\mstarE{1.48}
\def\mstareE{0.10}
\def\bvE{0.95}

\def\vmagE{8.13}

\def\mvE{2.6}
\def\mveE{0.1}
\def\vsiniE{2.97}
\def\ageE{3.0}
\def\ageeE{0.4}
\def\rstarE{4.2}
\def\rstareE{0.1}
\def\lstarE{10.5}
\def\lstareE{0.5}
\def\teffE{5052}
\def\loggE{3.5}
\def\feE{+0.05}

\def\dE{126}
\def\deE{11}
\def\starF{HD\,102329}
\def\pF{778.1}
\def\peF{7.1}

\def\tpF{15096}
\def\tpeF{20}
\def\eF{0.211}
\def\eeF{0.044}

\def\omF{182}
\def\omeF{10}
\def\kF{84.8}
\def\keF{3.4}
\def\jittF{6.8}
\def\jitteF{0.9}
\def\msiniF{5.9}
\def\msinieF{0.3}
\def\arelF{2.01}
\def\areleF{0.05}
\def\rmsF{7.2}

\def\nobsF{20}
\def\mstarF{1.95}
\def\mstareF{0.14}
\def\bvF{1.04}

\def\vmagF{8.04}

\def\mvF{2.1}
\def\mveF{0.3}
\def\vsiniF{2.60}
\def\ageF{1.6}
\def\ageeF{0.4}
\def\rstarF{6.3}
\def\rstareF{0.1}
\def\lstarF{19.6}
\def\lstareF{0.5}
\def\teffF{4830}
\def\loggF{3.0}
\def\feF{+0.30}

\def\dF{158}
\def\deF{21}
\def\starG{HD\,108863}
\def\pG{443.4}
\def\peG{4.2}

\def\tpG{15516}
\def\tpeG{70}

\def\eupG{0.10}
\def\omG{177}
\def\omeG{60}
\def\kG{45.2}
\def\keG{1.7}
\def\jittG{4.9}
\def\jitteG{0.9}
\def\msiniG{2.6}
\def\msinieG{0.2}
\def\arelG{1.40}
\def\areleG{0.03}
\def\rmsG{5.1}

\def\nobsG{24}
\def\mstarG{1.85}
\def\mstareG{0.13}
\def\bvG{0.99}

\def\vmagG{7.89}

\def\mvG{2.2}
\def\mveG{0.2}
\def\vsiniG{1.06}
\def\ageG{1.8}
\def\ageeG{0.4}
\def\rstarG{5.6}
\def\rstareG{0.1}
\def\lstarG{16.8}
\def\lstareG{0.5}
\def\teffG{4956}
\def\loggG{3.2}
\def\feG{+0.20}

\def\dG{139}
\def\deG{15}
\def\starH{HD\,116029}
\def\pH{670.2}
\def\peH{8.3}

\def\tpH{15220}
\def\tpeH{200}

\def\eupH{0.21}
\def\omH{360}
\def\omeH{100}
\def\kH{36.6}
\def\keH{3.6}
\def\jittH{6.8}
\def\jitteH{0.9}
\def\msiniH{2.1}
\def\msinieH{0.2}
\def\arelH{1.73}
\def\areleH{0.04}
\def\rmsH{6.9}

\def\nobsH{21}
\def\mstarH{1.58}
\def\mstareH{0.11}
\def\bvH{1.009}

\def\vmagH{8.04}

\def\mvH{2.6}
\def\mveH{0.2}
\def\vsiniH{0.46}
\def\ageH{2.7}
\def\ageeH{0.5}
\def\rstarH{4.6}
\def\rstareH{0.1}
\def\lstarH{11.3}
\def\lstareH{0.5}
\def\teffH{4951}
\def\loggH{3.4}
\def\feH{+0.18}

\def\dH{123.2}
\def\deH{9.9}
\def\starI{HD\,131496}
\def\pI{883}
\def\peI{29}

\def\tpI{16040}
\def\tpeI{100}
\def\eI{0.163}
\def\eeI{0.073}

\def\omI{22}
\def\omeI{40}
\def\kI{35.0}
\def\keI{2.1}
\def\jittI{6.8}
\def\jitteI{0.8}
\def\msiniI{2.2}
\def\msinieI{0.2}
\def\arelI{2.09}
\def\areleI{0.07}
\def\rmsI{6.3}

\def\nobsI{43}
\def\mstarI{1.61}
\def\mstareI{0.11}
\def\bvI{1.04}

\def\vmagI{7.96}

\def\mvI{2.8}
\def\mveI{0.2}
\def\vsiniI{0.48}
\def\ageI{2.7}
\def\ageeI{0.5}
\def\rstarI{4.3}
\def\rstareI{0.1}
\def\lstarI{9.8}
\def\lstareI{0.5}
\def\teffI{4927}
\def\loggI{3.3}
\def\feI{+0.25}

\def\dI{110.0}
\def\deI{9.4}
\def\starJ{HD\,158038}
\def\pJ{521.0}
\def\peJ{6.9}

\def\tpJ{15491}
\def\tpeJ{20}
\def\eJ{0.291}
\def\eeJ{0.093}

\def\omJ{334}
\def\omeJ{10}
\def\kJ{33.9}
\def\keJ{3.3}
\def\trendJ{63.5}
\def\trendeJ{1.5}

\def\jittJ{6.1}
\def\jitteJ{0.9}
\def\msiniJ{1.8}
\def\msinieJ{0.2}
\def\arelJ{1.52}
\def\areleJ{0.04}
\def\rmsJ{4.7}

\def\nobsJ{24}
\def\mstarJ{1.65}
\def\mstareJ{0.12}
\def\bvJ{1.04}

\def\vmagJ{7.64}

\def\mvJ{2.6}
\def\mveJ{0.1}
\def\vsiniJ{1.66}
\def\ageJ{2.5}
\def\ageeJ{0.3}
\def\rstarJ{4.8}
\def\rstareJ{0.1}
\def\lstarJ{11.9}
\def\lstareJ{0.5}
\def\teffJ{4897}
\def\loggJ{3.2}
\def\feJ{+0.28}

\def\dJ{103.6}
\def\deJ{7.9}
\def\starK{HD\,99706}
\def\pK{868}
\def\peK{31}

\def\tpK{15219}
\def\tpeK{30}
\def\eK{0.365}
\def\eeK{0.10}

\def\omK{359}
\def\omeK{20}
\def\kK{22.4}
\def\keK{2.2}
\def\trendK{-7.4}
\def\trendeK{2.0}

\def\jittK{4.6}
\def\jitteK{0.9}
\def\msiniK{1.4}
\def\msinieK{0.1}
\def\arelK{2.14}
\def\areleK{0.08}
\def\rmsK{3.7}

\def\nobsK{24}
\def\mstarK{1.72}
\def\mstareK{0.12}
\def\bvK{1.0}

\def\vmagK{7.81}

\def\mvK{2.3}
\def\mveK{0.2}
\def\vsiniK{0.89}
\def\ageK{2.1}
\def\ageeK{0.4}
\def\rstarK{5.4}
\def\rstareK{0.1}
\def\lstarK{15.4}
\def\lstareK{0.5}
\def\teffK{4932}
\def\loggK{3.2}
\def\feK{+0.14}

\def\dK{129}
\def\deK{11}
\def\starL{HD\,30856}
\def\pL{912}
\def\peL{41}

\def\tpL{15260}
\def\tpeL{150}

\def\eupL{0.24}
\def\omL{180}
\def\omeL{60}
\def\kL{31.9}
\def\keL{2.7}
\def\jittL{6}
\def\jitteL{1}
\def\msiniL{1.8}
\def\msinieL{0.2}
\def\arelL{2.00}
\def\areleL{0.08}
\def\rmsL{5.2}

\def\nobsL{16}
\def\mstarL{1.35}
\def\mstareL{0.094}
\def\bvL{0.961}

\def\vmagL{8.07}

\def\mvL{2.7}
\def\mveL{0.2}
\def\vsiniL{2.85}
\def\ageL{3.8}
\def\ageeL{1}
\def\rstarL{4.2}
\def\rstareL{0.1}
\def\lstarL{9.9}
\def\lstareL{0.5}
\def\teffL{4982}
\def\loggL{3.4}
\def\feL{-0.06}

\def\dL{118.1}
\def\deL{9.9}
\def\starM{HD\,82886}
\def\pM{705}
\def\peM{34}

\def\tpM{15200}
\def\tpeM{160}

\def\eupM{0.27}
\def\omM{347}
\def\omeM{80}
\def\kM{28.7}
\def\keM{2.1}
\def\trendM{7.5}
\def\trendeM{2.3}

\def\jittM{7.3}
\def\jitteM{0.9}
\def\msiniM{1.3}
\def\msinieM{0.1}
\def\arelM{1.65}
\def\areleM{0.06}
\def\rmsM{7.7}

\def\nobsM{28}
\def\mstarM{1.06}
\def\mstareM{0.074}
\def\bvM{0.864}

\def\vmagM{7.78}

\def\mvM{2.3}
\def\mveM{0.1}
\def\vsiniM{0.43}
\def\ageM{7}
\def\ageeM{2}
\def\rstarM{4.8}
\def\rstareM{0.1}
\def\lstarM{13.9}
\def\lstareM{0.5}
\def\teffM{5112}
\def\loggM{3.4}
\def\feM{-0.31}

\def\dM{125}
\def\deM{12}
\def\starN{HD\,96063}
\def\pN{361.1}
\def\peN{9.9}

\def\tpN{15260}
\def\tpeN{120}

\def\eupN{0.28}
\def\omN{90}
\def\omeN{100}
\def\kN{25.9}
\def\keN{3.5}
\def\jittN{6}
\def\jitteN{1}
\def\msiniN{0.9}
\def\msinieN{0.1}
\def\arelN{0.99}
\def\areleN{0.03}
\def\rmsN{5.4}

\def\nobsN{15}
\def\mstarN{1.02}
\def\mstareN{0.072}
\def\bvN{0.86}

\def\vmagN{8.37}

\def\mvN{2.4}
\def\mveN{0.3}
\def\vsiniN{0.87}
\def\ageN{9}
\def\ageeN{3}
\def\rstarN{4.5}
\def\rstareN{0.1}
\def\lstarN{12.7}
\def\lstareN{0.5}
\def\teffN{5148}
\def\loggN{3.6}
\def\feN{-0.30}

\def\dN{158}
\def\deN{20}
\def\starO{HD\,98219}
\def\pO{436.9}
\def\peO{4.5}

\def\tpO{15140}
\def\tpeO{40}

\def\omO{47}
\def\omeO{30}
\def\kO{41.2}
\def\keO{1.9}
\def\jittO{4}
\def\jitteO{1}
\def\msiniO{1.8}
\def\msinieO{0.1}
\def\arelO{1.23}
\def\areleO{0.03}
\def\rmsO{3.6}

\def\nobsO{17}
\def\mstarO{1.30}
\def\mstareO{0.091}
\def\bvO{0.96}

\def\vmagO{8.21}

\def\mvO{2.6}
\def\mveO{0.2}
\def\vsiniO{0.30}
\def\ageO{4}
\def\ageeO{1}
\def\rstarO{4.5}
\def\rstareO{0.1}
\def\lstarO{11.2}
\def\lstareO{0.5}
\def\teffO{4992}
\def\loggO{3.5}
\def\feO{-0.02}

\def\dO{134}
\def\deO{12}
\def\starP{HD\,106270}
\def\pP{2890}
\def\peP{390}

\def\tpP{14830}
\def\tpeP{390}
\def\eP{0.402}
\def\eeP{0.054}

\def\omP{15.4}
\def\omeP{4}
\def\kP{142.1}
\def\keP{6.9}
\def\jittP{7.5}
\def\jitteP{0.9}
\def\msiniP{11.0}
\def\msinieP{0.8}
\def\arelP{4.3}
\def\areleP{0.4}
\def\rmsP{8.4}

\def\nobsP{20}
\def\mstarP{1.32}
\def\mstareP{0.092}
\def\bvP{0.74}

\def\vmagP{7.73}

\def\mvP{3.1}
\def\mveP{0.2}
\def\vsiniP{3.13}
\def\ageP{4.3}
\def\ageeP{0.6}
\def\rstarP{2.5}
\def\rstareP{0.1}
\def\lstarP{5.7}
\def\lstareP{0.5}
\def\teffP{5638}
\def\loggP{3.9}
\def\feP{+0.08}

\def\dP{84.9}
\def\deP{5.7}
\def\starQ{HD\,152581}
\def\pQ{689}
\def\peQ{13}

\def\tpQ{15320}
\def\tpeQ{190}

\def\eupQ{0.22}
\def\omQ{321}
\def\omeQ{90}
\def\kQ{36.6}
\def\keQ{1.8}
\def\jittQ{5.5}
\def\jitteQ{0.9}
\def\msiniQ{1.5}
\def\msinieQ{0.1}
\def\arelQ{1.48}
\def\areleQ{0.04}
\def\rmsQ{4.7}

\def\nobsQ{24}
\def\mstarQ{0.927}
\def\mstareQ{0.065}
\def\bvQ{0.90}

\def\vmagQ{8.54}

\def\mvQ{2.2}
\def\mveQ{0.4}
\def\vsiniQ{0.50}
\def\ageQ{12}
\def\ageeQ{3}
\def\rstarQ{4.8}
\def\rstareQ{0.1}
\def\lstarQ{14.9}
\def\lstareQ{0.6}
\def\teffQ{5155}
\def\loggQ{3.4}
\def\feQ{-0.46}

\def\dQ{186}
\def\deQ{33}
\def\starR{HD\,1502}
\def\pR{431.8}
\def\peR{3.1}

\def\tpR{15227}
\def\tpeR{20}
\def\eR{0.101}
\def\eeR{0.036}

\def\omR{219}
\def\omeR{20}
\def\kR{60.7}
\def\keR{1.9}
\def\jittR{9.3}
\def\jitteR{0.7}
\def\msiniR{3.1}
\def\msinieR{0.2}
\def\arelR{1.31}
\def\areleR{0.03}
\def\rmsR{10.9}

\def\nobsR{51}
\def\mstarR{1.61}
\def\mstareR{0.11}
\def\bvR{0.92}

\def\vmagR{8.52}

\def\mvR{2.5}
\def\mveR{0.3}
\def\vsiniR{2.70}
\def\ageR{2.4}
\def\ageeR{0.5}
\def\rstarR{4.5}
\def\rstareR{0.1}
\def\lstarR{11.6}
\def\lstareR{0.5}
\def\teffR{5049}
\def\loggR{3.4}
\def\feR{0.09}

\def\dR{159}
\def\deR{19}
\def\starS{HD\,142245}
\def\pS{1299}
\def\peS{48}

\def\tpS{14760}
\def\tpeS{240}

\def\eupS{0.32}
\def\omS{234}
\def\omeS{60}
\def\kS{24.8}
\def\keS{2.6}
\def\jittS{5.5}
\def\jitteS{0.9}
\def\msiniS{1.9}
\def\msinieS{0.2}
\def\arelS{2.77}
\def\areleS{0.09}
\def\rmsS{4.8}

\def\nobsS{19}
\def\mstarS{1.69}
\def\mstareS{0.12}
\def\bvS{1.04}

\def\vmagS{7.63}

\def\mvS{2.4}
\def\mveS{0.1}
\def\vsiniS{2.66}
\def\ageS{2.3}
\def\ageeS{0.3}
\def\rstarS{5.2}
\def\rstareS{0.1}
\def\lstarS{13.5}
\def\lstareS{0.5}
\def\teffS{4878}
\def\loggS{3.3}
\def\feS{+0.23}

\def\dS{109.5}
\def\deS{7.4}

\def\sB{0.108}

\def\sC{0.133}

\def\sD{0.130}

\def\sE{0.140}

\def\sF{0.129}

\def\sG{0.127}

\def\sH{0.133}

\def\sI{0.121}

\def\sJ{0.119}

\def\sK{0.132}

\def\sL{0.130}

\def\sM{0.135}

\def\sN{0.146}

\def\sO{0.136}

\def\sP{0.186}

\def\sQ{0.146}

\def\sR{0.146}

\def\sS{0.122}

\def\nplup{EIGHTEEN}
\def\npllet{eighteen}

\def\npl{18}

\begin{document}
\title{Retired A Stars and Their Companions VII. \\ \nplup\ New Jovian Planets$^1$}  

\author{John Asher Johnson\altaffilmark{2,9},
Christian Clanton\altaffilmark{2,9},
Andrew W. Howard\altaffilmark{4},
Brendan P. Bowler\altaffilmark{3},
Gregory W. Henry\altaffilmark{5},
Geoffrey W. Marcy\altaffilmark{4},
Justin R. Crepp\altaffilmark{2},
Michael Endl\altaffilmark{6},
William D. Cochran\altaffilmark{6},
Phillip J. MacQueen\altaffilmark{6},
Jason T. Wright\altaffilmark{7,8},
Howard Isaacson\altaffilmark{3}
}

\email{johnjohn@astro.caltech.edu}

\altaffiltext{1}{ Based on observations obtained at the
W.M. Keck Observatory, McDonald Observatory and the 
Hobby-Ebberly Telescope. Keck is operated jointly by the
University of California and the California Institute of
Technology. Keck time has been granted by Caltech, the University of
Hawaii, NASA and the University of California.}
\altaffiltext{2}{Department of Astrophysics,
  California Institute of Technology, MC 249-17, Pasadena, CA 91125}
\altaffiltext{3}{Institute for Astronomy, University of Hawai'i, 2680
  Woodlawn Drive, Honolulu, HI 96822} 
\altaffiltext{4}{Department of Astronomy, University of California,
Mail Code 3411, Berkeley, CA 94720}
\altaffiltext{5}{Center of Excellence in Information Systems, Tennessee
  State University, 3500 John A. Merritt Blvd., Box 9501, Nashville, TN 37209}
\altaffiltext{6}{McDonald Observatory, University of Texas at Austin,
  TX, 78712-0259, USA}
\altaffiltext{7}{Department of Astronomy \& Astrophysics, The
  Pennsylvania State University, University Park,  PA 16802}
\altaffiltext{8}{Center for Exoplanets and Habitable Worlds, The
  Pennsylvania State University, University Park, PA 16802}
\altaffiltext{9}{NASA Exoplanet Science Institute (NExScI), CIT Mail
  Code 100-22, 770 South Wilson Avenue, Pasadena, CA 91125}

\begin{abstract}
We report the detection of \npllet\ Jovian planets discovered as part
of our Doppler survey of subgiant stars at Keck Observatory, with
follow-up Doppler and photometric observations made at McDonald and
Fairborn Observatories, respectively. The host stars have masses
$\mstarQ \leq M_\star/M_\odot \leq \mstarF$, radii 
$\rstarP \leq R_\star/R_\odot \leq \rstarB$, and metallicities $\feQ
\leq$~\feh~$ \leq \feF$. The planets have minimum masses 
0.9~\mjup~$ \leq$ \msini $\lesssim 13$~\mjup\ and semimajor axes 
$a \geq \arelB$~AU. These detections represent a 50\% increase in the
number of planets known to orbit stars more massive than
1.5~\msun\ and provide valuable additional information about the
properties of planets around stars more massive than the Sun. 
\end{abstract}


\section{Introduction}

Jupiter-mass planets are not uniformly distributed around all stars in
the galaxy. Rather, the rate of planet occurrence is 
intimately tied to the physical properties of the the stars they
orbit \citep{johnson10c,howard11,sl11}. Radial velocity surveys have
demonstrated that the likelihood 
that a star harbors a giant planet with a minimum mass \msini~$\gtrsim
0.5$~\mjup\ increases with both stellar 
metallicity and mass\footnote{Some studies indicate a lack of a
planet-metallicity relationship among planet-hosting K-giants
\citep{pasquini07,sato08b}. However, a planet-metallicity
correlation is evident among subgiants, which probe an overlapping
range of stellar masses and convective envelope depths
\citep{fischer05b,johnson10c,ghezzi10}.}   
\citep{gonzalez97,santos04,fischer05b,johnson10c,sl10,brugamyer11}. This    
result has both informed models of giant planet formation
\citep{ida04,laughlin04,thommes08,kennedy08,mordasini09} and
pointed the way toward additional exoplanet discoveries
\citep{laughlin00,marois08}. 

The increased abundance 
of giant planets around massive, metal-rich stars may be a reflection
of their more massive, dust-enriched circumstellar disks, which form
protoplanetary cores more efficiently
\citep{ida04,fischer05b,thommes06,wyatt07}. In 
the search for additional planets in the 
Solar neighborhood, metallicity-biased Doppler surveys have greatly
increased the number of close-in, transiting exoplanets around nearby,
bright stars, thereby enabling detailed studies of exoplanet
atmospheres \citep{fischer05a,dasilva06,charbonneau07}. Similarly, future 
high-contrast imaging surveys will 
likely benefit from enriching their target lists with
intermediate-mass A- and F-type stars \citep{marois08,crepp11}. 

Occurrence rate is not the only aspect of exoplanets that correlates with
stellar mass. Just when exoplanet researchers were growing accustomed
to short-period and highly eccentric planets around Sun-like stars,
surveys of evolved stars revealed that the orbital properties of
planets are very different at higher stellar masses. Stars more
massive than 1.5~\msun\ may have a higher overall occurrence of
Jupiters than do 
Sun-like stars, but they exhibit a marked paucity of planets with
semimajor axes $a \lesssim 1$~AU \citep{johnson07,sato08a}. This is not an
observational bias since close-in, giant planets produce readily
detectable Doppler signals. There is    
also growing evidence that planets around more massive stars tend to
have larger minimum masses \citep{lovis07,bowler10}, and occupy less
eccentric orbits compared to planets around
Sun-like stars \citep{johnson08c}.  

M-type dwarfs also exhibit a deficit of ``hot Jupiters,''
albeit with a lower overall occurrence of giant planets at all periods 
\citep{endl03, johnson10c}. However, a recent analysis of the
transiting planets 
detected by the spaced-based \emph{Kepler} mission shows that the
occurrence of close-in, \emph{low-mass} planets 
($P < 50$~days, $M_P \lesssim 0.1$~\mjup) increases steadily with
\emph{decreasing} stellar mass \citep{howard11}. Also counter to the
statistics of Jovian planets, low-mass planets are found quite
frequently around low-metallicity stars
\citep{sousa08,valenti09}. These results strongly suggest that stellar
mass is a key variable in the formation and subsequent orbital
evolution of planets, and that the formation of gas giants is likely a
threshold process that leaves behind a multitude of ``failed cores''
with masses of order $10$~\mearth.  

To study the properties of planets around stars more massive than the
Sun, we are conducting a Doppler survey of intermediate-mass subgiant
stars, also known as the ``retired'' A-type stars
\citep{johnson06b}. Main-sequence stars 
with masses greater than $\approx 1.3$~\msun\ (spectral types
$\lesssim$\,F8) are challenging targets for Doppler surveys because
they are hot and rapidly rotating \citep[$T_{\rm eff} > 6300$,
\vsini$\gtrsim 30$~\ks;][]{galland05}. However, post-main-sequence
stars located on 
the giant and subgiant branches are cooler and have much slower
rotation rates than their main-sequence cohort. Their spectra therefore
exhibit a higher density of narrow absorption lines that are ideal for
precise Doppler-shift measurements. 

Our survey has resulted in the detection 
of 16 planets around 14 intermediate-mass ($M_\star \gtrsim
1.5$~\msun) stars, including two multiplanet systems, the first
Doppler-detected hot Jupiter around an intermediate-mass star, and 4  
additional Jovian planets around less massive 
subgiants \citep[][]{johnson06b,
  johnson07,johnson08b,bowler10, peek09,
  johnson10b,johnson10d,johnson11a}. In this 
contribution we announce the detection of \npl\ new giant 
exoplanets orbiting subgiants spanning a wide range of
stellar physical properties.

\section{Observations and Analysis}
\label{sec:obs}

\subsection{Target Stars}

\begin{figure}
\epsscale{1.1}
\plotone{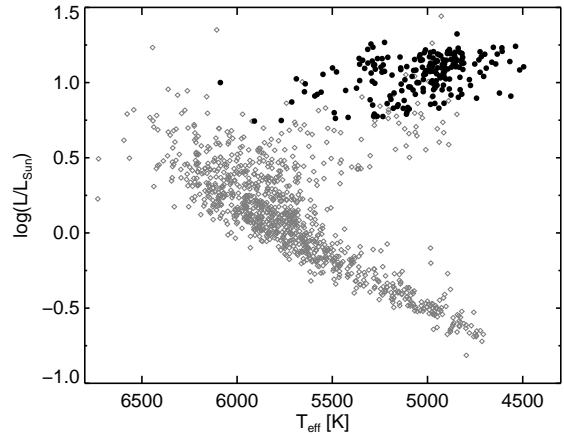}
\caption{Distribution of the effective temperatures and luminosities
  of the Keck sample of subgiants (filled circles) compared to the
  full CPS Keck target sample (gray diamonds). 
  \label{fig:hr}}
\end{figure}

The details of the target selection of our Doppler survey of evolved
stars at Keck Observatory have been described in detail by
\citet[e.g.][]{johnson06b,peek09,johnson10b}. In summary, we have
selected subgiants from the \emph{Hipparcos} catalog \citep{hipp2} 
based on $B-V$
colors and absolute magnitudes $M_V$ so as to avoid K-type giants that
are observed as part of other Doppler surveys 
\citep[e.g.][]{hatzes03,sato05,reffert06}, and exhibit jitter
levels in excess of 10~\ms\ \citep{hekker06}. We also selected stars
in a region of the temperature-luminosity plane in which stellar model
grids of various masses are well separated and
correspond to masses $M_\star > 1.3$~\msun\ at Solar
metallicity according to the \citet{girardi02} model grids. However,
some of our stars have sub-Solar metallicities 
([Fe/H]~$<0$) and correspondingly lower masses down to $\approx
1$~\msun. Our sample of 240 subgiants monitored at Keck Observatory
(excluding the Lick Observatory sample described by
\citet{johnson06b}) is 
shown in Figure~\ref{fig:hr} and  
compared to the full target sample of the California Planet Survey
(CPS).

\subsection{Spectra and Doppler-Shift Measurements}

We obtained spectroscopic observations of our sample of subgiants at
Keck Observatory using the 
HIRES spectrometer with a resolution of $R \approx 55,000$ with the
B5 decker (0\farcs86 width) and red cross-disperser \citep{vogt94}.
We use the HIRES 
exposure meter to ensure that all observations receive uniform flux
levels independent of atmospheric transparency variations, and to
provide the photon-weighted exposure midpoint which is used for the
barycentric correction. Under nominal atmospheric conditions, a $V=8$
target requires an exposure time of 90 seconds and results in a
signal-to-noise ratio of 190 at 5800\,\AA\ for our sample comprising
mostly early K-type stars.   

Normal program observations are made through a
temperature-controlled Pyrex cell containing gaseous iodine, which is
placed just in front of the entrance slit of the spectrometer. The
dense set of narrow molecular lines imprinted on each stellar spectrum
from 5000~\AA\ to 
6200~\AA\ provides a robust, simultaneous wavelength calibration for
each observation, as well as information about the shape of the
spectrometer's instrumental response \citep{marcy92b}. Radial
velocities (RV) are measured with respect to an iodine-free
``template'' observation that has had the HIRES instrumental profile
removed through deconvolution. Differential Doppler shifts 
are measured from each spectrum using the forward-modeling procedure
described by \citet{butler96}, with subsequent improvements over the
years by the CPS team \citep[e.g.][]{howard11b}. The instrumental
uncertainty of each 
measurement is estimated based on the weighted standard deviation of
the mean 
Doppler-shift measured from each of $\approx700$ independent
2--\AA\ spectral regions. In a few instances we made two or more
successive observations 
of the same star and averaged the velocities in 2\,hour time intervals,
thereby reducing the associated measurement uncertainty.  

We have also obtained additional spectra for \starR\ in
collaboration with the McDonald Observatory planet search team. A 
total of 54 RV measurements were collected for HD~1502: 32 with the
2.7\,m Harlan J. Smith Telescope (HJST) and its Tull Coude
Spectrograph \citep{tull95}, and 22 
with the High Resolution Spectrograph 
\citep[HRS;][]{tull98} at the Hobby-Eberly Telescope
\citep[HET;][]{ramsey98}. On each spectrometer we use a
sealed and temperature 
controlled iodine cell as velocity metric and to allow PSF
reconstruction. The spectral resolving power for the HRS and Tull
spectrograph is set 
to $R=60,000$. Precise differential RVs are computed using the {\it 
Austral} I$_2$-data modeling algorithm \citep{endl00}. 

The RV measurements are listed in
Tables\,\ref{vel1502}--\ref{vel158038} together with the 
Heliocentric Julian Date of observation (HJD) and internal measurement
uncertainties, excluding the jitter contribution described in the
\S~\ref{sec:orbit}.

\subsection{Stellar Properties}
\label{sec:stellar_properties}

\begin{figure*}[!t]
\epsscale{1.1}
\plotone{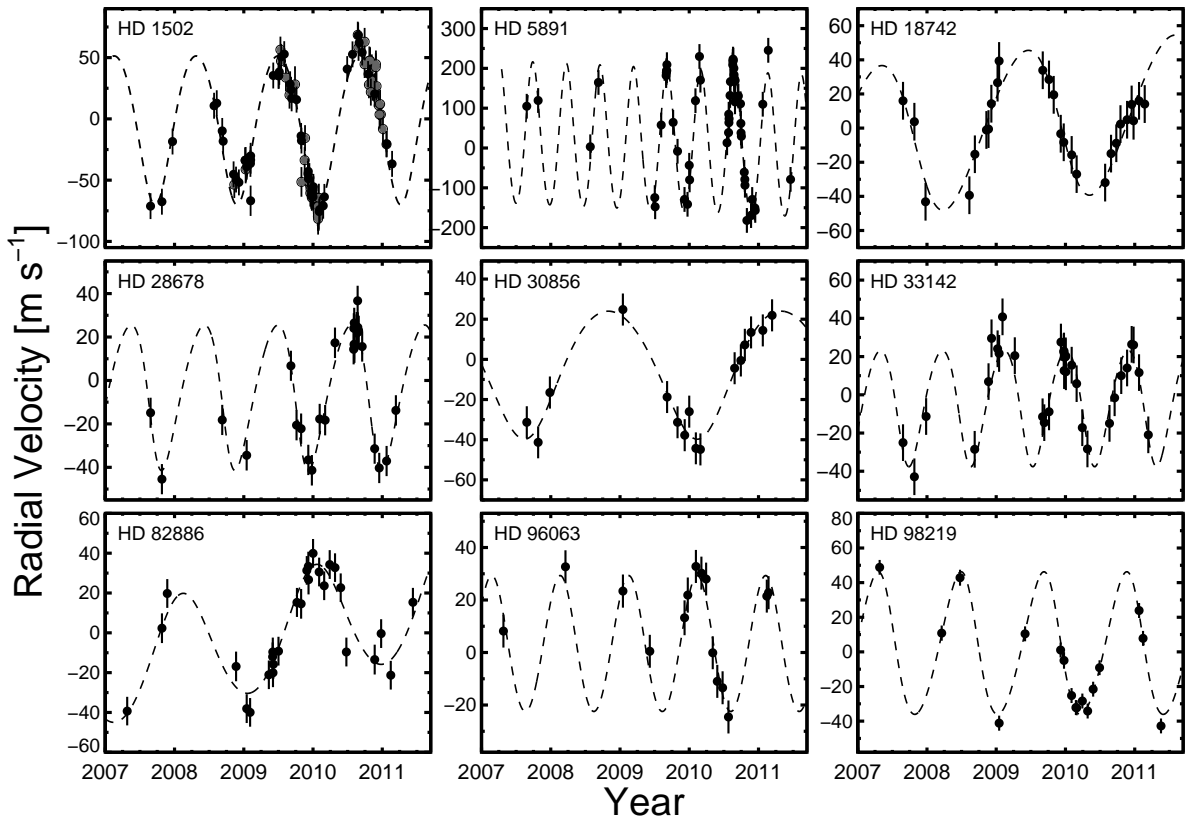}
\caption{Relative RVs of 9 stars measured at Keck Observatory. The error
  bars are the quadrature sum of the internal measurement uncertainties
  and jitter estimates. The dashed line shows the best-fitting orbit
  solution of a single Keplerian orbit, with a linear trend where
  appropriate. 
  \label{fig:nine_orbits1}}
\end{figure*}

\begin{figure*}[!t]
\epsscale{1.1}
\plotone{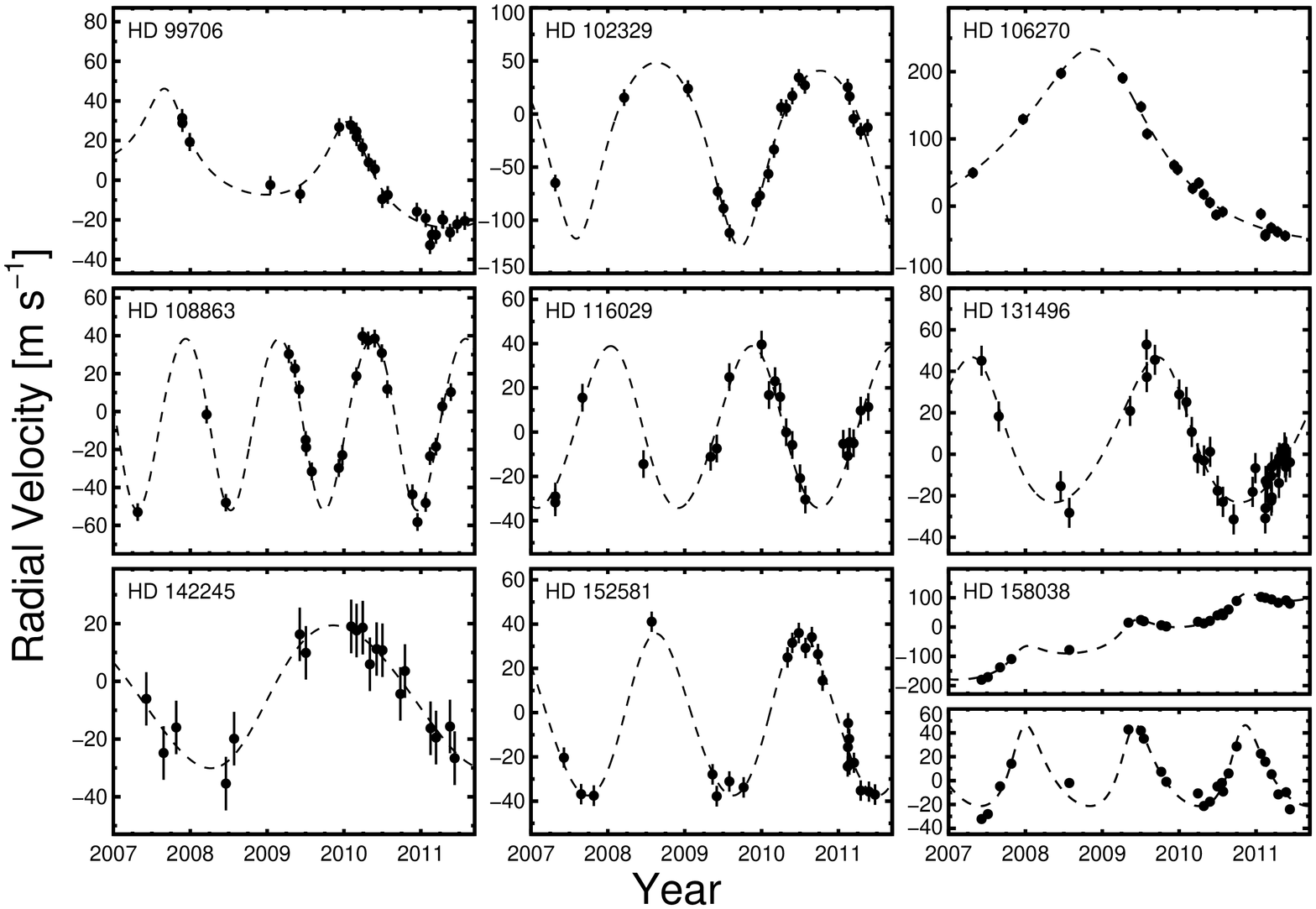}
\caption{Relative RVs of 9 stars measured at Keck Observatory. The error
  bars are the quadrature sum of the internal measurement uncertainties
  and jitter estimates. The dashed line shows the best-fitting orbit
  solution of a single Keplerian orbit. The split, lower-right panel
  shows the   orbit of HD\,158038  with a linear trend ({\it top}) and
  with the trend removed ({\it bottom}). 
  \label{fig:nine_orbits2}}
\end{figure*}

We use the Iodine-free template spectra to estimate
atmospheric 
parameters of the target stars with the LTE spectroscopic analysis 
package {\it Spectroscopy Made Easy} \citep[SME;][]{valenti96}, 
as described by \citet{valenti05} and \citet{fischer05b}. 
Subgiants have lower surface gravities than dwarfs, and the damping
wings of the Mg\,I\,b triplet lines therefore provide weaker constraints
on the surface gravity, which is in turn degenerate with 
effective temperature and metallicity. To constrain $\log{g}$ we
use the iterative scheme of \citet{valenti09}, which ties the
SME-derived value of $\log{g}$ to the gravity inferred from
interpolating the stellar luminosity, temperature and metallicity onto
the  
Yonsei-Yale \citep[Y$^2$;][]{y2} stellar model grids, which also give
the stellar age and mass. The model-based $\log{g}$ is held fixed in
a second SME analysis, and the process is iterated until convergence
is met between the model-based and spectroscopically measured surface
gravity, which 
results in best-fitting estimates of \teff, \logg, \feh, and \vsini. 

We perform our model-grid interpolations using a Bayesian framework
similar to that described by \citet{takeda08}. We incorporate prior
constraints on the stellar mass based on the stellar initial mass function 
and the differential evolutionary timescales of stars in various
regions of the theoretical H--R diagram. These priors tend
to decrease the stellar mass inferred for a star of a given effective
temperature, luminosity and metallicity compared to a naive
interpolation onto the stellar model grids \citep{lloyd11}. 

We determine the luminosity of each star from the apparent V-band 
magnitude and 
parallax from {\it Hipparcos} \citep{hipp2}, and the bolometric
correction based the effective temperature relationship given by
\citet{vandenberg03}\footnote{Previous papers in this series
  \citep[e.g.][]{johnson10b,johnson11a} incorrectly cited use of the
  \citet{flower96} bolometric corrections.}. Stellar radii are
estimated using the 
Stefan-Boltzmann relationship and the measured $L_\star$ and $T_{\rm eff}$. 
We also measure the chromospheric emission in the \caii\ line cores 
\citep{wright04b, isaacson10}, providing an $S_\mathrm{HK}$ value on the
Mt. Wilson system.

The stellar properties of the \npllet\ stars presented herein 
are summarized in Table \ref{tab:stars}.

\subsection{Photometric Measurements}

We acquired photometric observations of 17 of the 18 planetary
candidate host stars  
with the T3 0.4~m automatic photometric telescope (APT) at Fairborn 
Observatory.  T3 observed each program star differentially with respect to 
two comparisons stars in the following sequence, termed a group observation: 
{\it K,S,C,V,C,V,C,V,C,S,K}, where $K$ is a check (or secondary comparison) 
star, $C$ is the primary comparison star, $V$ is the target star, and $S$ 
is a sky reading.  Three $V-C$ and two $K-C$ differential magnitudes are 
computed from each sequence and averaged to create group means.  Group mean 
differential magnitudes with internal standard deviations greater than 
0.01 mag were rejected to eliminate the observations taken under 
non-photometric conditions.  The surviving group means were corrected for 
differential extinction with nightly extinction coefficients, transformed 
to the Johnson system with yearly-mean transformation coefficients, and 
treated as single observations thereafter.  The precision of a single 
group-mean observation is usually in the range $\sim$0.003--0.006 mag 
\citep[e.g.,][]{henry00c}, depending on the brightness of the
stars within the group, the quality of the night, and the airmass of the
observation.  Further information on the operation of the T3 APT can be 
found in \citet{henry95,henry95b} and \citet{eaton03}.
 
Our photometric observations are useful for eliminating potential false 
positives from our sample of new planets. For example,
\citet{queloz01b} and  
\citet{paulson04} have demonstrated how rotational modulation in the 
visibility of starspots on active stars can result in periodic radial 
velocity variations and, therefore, the potential for erroneous planetary 
detections.  Photometric results for the 17 stars in the present sample 
are given in Table~\ref{tab:phot}.  Columns 7--10 give the standard
deviations of the  
$V-C$ and $K-C$ differential magnitudes in the $V$ and $B$ passbands 
with $3\sigma$ outliers removed.  All of the standard deviations are 
small and consistent with the measurement precision of the telescope.  
Periodogram analysis of each data set found no significant periodicity
between 1 and 100 days.  

We conclude that all 17 planetary condidate stars in
Table~\ref{tab:phot}, as well as  
all of their comparison and check stars, are constant to the limit of our 
photometric precision.  The lack of evidence for photometric variability 
provides support for the planetary interpretation of the radial velocity 
variations.

Although we do not have photometric measurements of HD 142245, we note
from Table 19 that HD 142245 has one of the lowest values for \shk\
in the sample.  Therefore, like the rest of the sample, HD 142245
should be photometrically stable.

\subsection{Orbit Analysis}
\label{sec:orbit}

As in \citet{johnson10b}, we perform a thorough search of the
RV time series of each star for the best-fitting Keplerian  orbital
model using the partially-linearized, least-squares 
fitting  procedure described in \citet{wrighthoward} and implemented
in the IDL package  
{\tt RVLIN}\footnote{http://exoplanets.org/code/}. 
The free parameters in our model are the velocity
semiamplitude $K$, period $P$, argument of periastron $\omega$, time
of periastron passage $T_p$, and the systemic velocity offset
$\gamma$. When fitting RVs from separate observatories we 
include additional offsets $\gamma_i$ for the different data sets. As
described in \S~\ref{sec:trends}, we also explore the existence of a
constant acceleration $\dot{\gamma}$ in each RV time series. 

In addition to the parameters describing the orbit, we also include 
an additional error contribution to our RV measurements due to stellar 
"jitter," which we denote by $s$. The jitter accounts for any
unmodeled noise sources 
intrinsic to the star such as rotational modulation of surface
inhomogeneities and pulsation \citep{saar98,
  wright05,makarov09,lagrange10}, and is added in quadrature to the
internal uncertainty of each RV measurement. 

Properly estimating the jitter contribution to the uncertainty of each
measurement is key to accurately estimating the confidence intervals for
each fitted parameter. Ignoring jitter will lead to underestimated
parameter uncertainties, rendering them less useful in future
statistical investigations of exoplanet properties. Similarly,
the equally common practice relying on a single value of the jitter
based on stars with properties similar to the target of interest
ignores variability in the jitter observed from star to star, and can
potentially overestimate the parameter uncertainties. For these
reasons we take the approach of allowing the jitter term to vary in our
orbit analyses, as described by e.g. \citet{ford07}. 

We estimate parameter uncertainties using a Markov-Chain Monte Carlo
(MCMC) algorithm \citep[See, e.g.][]{ford05,winn07}. MCMC is a
Bayesian inference technique that uses the data together with prior
knowledge to explore the shape of the posterior probability density
function (PDF) for  each parameter of an input model. MCMC with the
Metropolis-Hastings  algorithm in particular provides an efficient
means of exploring high-dimensional parameter  space and mapping out
the posterior PDF for each model parameter.  

At each chain link in
our MCMC analysis, one parameter is selected at random and is altered
by drawing a random variate from a transition probability
distribution.  If the resulting value of the likelihood $\mathcal{L}$ 
for the trial orbit is greater than the previous value, then the trial
orbital parameters are accepted and added to the chain.  If not, then
the probability of adopting the new value is set by the ratio of the
probabilities from the previous and current trial steps.  If the
current trial is rejected then the parameters from the previous step
are adopted.  The size of the transition function determines the
effiency of convergence. If it is too narrow then the full exploration
of parameter space is slow and the chain is susceptible local minima;
if it is too broad then the chain exhibits large jumps and the
acceptance rates are low. 

Rather than minimizing
\chis, we maximize the logarithm of the likelihood of the data, given by

\begin{equation}
\ln{\mathcal{L}} = -\sum_{i=1}^{N_{\rm obs}} \ln{\sqrt{2\pi (\sigma_i + s)^2}}
- \frac{1}{2} \sum_{i=1}^{N_{\rm obs}} \left[\frac{v_i - v_{\rm
    m}(t_i)}{\sigma_i + s}\right]^2
\label{eqn:logl}
\end{equation}

\noindent where $v_i$ and $\sigma_i$ are the $i$th velocity
measurement and its associated measurement error; $v_{\rm
  m}(t_i)$ is the Keplerian model at time $t_i$;  $s$ is the
jitter; and the sum is performed over all $N_{\rm obs}$
measurements. If $s=0$, then the first
term on the right side---the normalization of the
probability---is a constant, and the second term becomes
$\frac{1}{2}\chi^2$. Thus, maximizing $\ln\mathcal{L}$ is
equivalent to minimizing $\chi^2$. Larger jitter values
more easily accomodate large deviations of the observed RV from the
model prediction,
but only under the penalty of a decreasing (more negative)
normalization term, which makes the overall likelihood smaller.

We impose uninformative priors for most of the free parameters (either
uniform or modified Jeffreys; e.g. \citep{gregory10}).  The notable
exception is jitter, for which we use a Gaussian prior with a mean of
5.1~\ms\ and a standard deviation of 1.5~\ms\ based on the distribution
of jitter values for a similar sample of  
intermediate-mass subgiants from \citet{johnson10b}.  

We use the best-fitting parameter values from {\tt RVLIN} as initial
guesses for our MCMC analysis. We choose normal transition probability
functions with constant (rather than adaptive) widths. The standard
deviations are iteratively chosen from a series of smaller chains so
that the acceptance  rates for each parameter are between 20\% and
30\%; each main chain is then run for $10^7$ steps.  The initial 10\%
of the chains are excluded from the final estimation of parameter
uncertainties to ensure uniform convergence.  We select the 15.9 and
84.1 percentile levels in the posterior distributions as the
``one-sigma'' confidence limits. In most cases the posterior
probability distributions were approximately Gaussian.

\subsection{Testing RV Trends}
\label{sec:trends}

To determine whether there is evidence for a linear velocity trend,
we use two separate methods: the Bayesian Information Criterion
\citep[BIC;][]{schwarz78, 
  liddle04}, and inspection of the MCMC posterior probability density
functions, as described by \citet{bowler10}. The BIC rewards
better-fitting models but penalizes overly complex models, and is
given by 

\begin{equation}
\mathrm{BIC} \equiv -2 \ln \mathcal{L_\mathrm{max}} + k \ln N,
\end{equation}

\noindent where $\mathcal{L_\mathrm{max}}$ is
the maximum likelihood 
for a particular model with $k$ free parameters and $N$ data
points. The relationship between
  $\mathcal{L_\mathrm{max}}$ and $\chi_{min}^2$ is only valid under
  the assumption that the RVs are normally distributed, which
  is approximately valid for our analyses. A difference of $\gtrsim$
2 between BIC values with and without a trend  indicates that there is
sufficient evidence for a more complex model \citep{kuha04}. 

We also  use the MCMC-derived probability density function (pdf) for
the velocity trend parameter  
to estimate the probability that a trend is actually present in the
data. We only adopt the model with the trend if the 99.7 percentile of
the pdf lies above or below 0~\ms~yr$^{-1}$.
The BIC and MCMC methods yield consistent results for the planet 
candidates presented in \S~\ref{results}, and in many cases the RV trend
is evident by visual inspection
of Figs.~\ref{fig:nine_orbits1}-\ref{fig:nine_orbits2}.

\section{Results}
\label{results}

We have detected \npllet\ new Jovian planets orbiting evolved,
subgiant stars. The RV time series of each host-star is plotted in
Figures~\ref{fig:nine_orbits1} and \ref{fig:nine_orbits2}, where the
error bars show the quadrature sum of the internal errors and the
jitter estimate as described in
\S~\ref{sec:orbit}. The RV measurements for each star are listed in
Tables~\ref{vel1502}--\ref{vel158038}, 
together with the Julian Date of observation and the internal
measurement uncertainties (without jitter). The best--fitting orbital
parameters and physical characteristics of the planets are summarized
in Table~\ref{tab:planets}, along with their
uncertainties. When appropriate we list notes for some of the
individual planetary systems. 

{\it \starC, \starD, \starM, \starK, \starJ}---The orbit models for
these 
stars include linear trends, which we interpret as additional
orbital companions with periods longer than the time baseline of the
observations. 

{\it \starN}---The period of this system is very close to 1 year,
raising the spectre that it may be an annual systematic error rather
than an actual planet. However, any such annual signal would most
likely be related to an error in the barycentric correction (BC), and
if present would cause the RVs to correlate with the BC. We checked and
found no such correlation between RV and BC. Further, we have never
seen an annual signal with an amplitude of this magnitude in any of
the several thousand targets monitored at Keck Observatory.

\begin{figure}
\epsscale{1.2}
\plotone{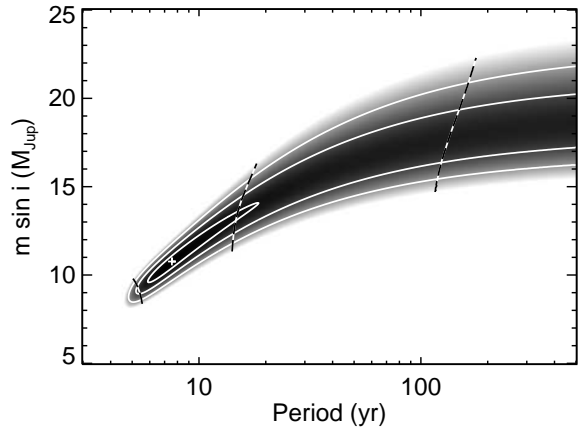}
\caption{Illustration of possible periods and minimum masses (\msini)
  for the companion orbiting \starP. At each value of \msini\ and $P$
  on the grid, the \emph{minimum} $\chi^2$ is shown in gray-scale. The
  solid contours denote the levels at which $\chi^2$ increases by 1, 4
  and 9 with respect to the minimum, from inside-out. The dashed
  contours denote   constant eccentricity values of $e = \{0.2, 0.6,
  0.9\}$ from left to 
  right. 
  \label{fig:MPcontour}}
\end{figure}

{\it \starP}---The reported orbit for this companion is long-period and we
only have limited phase coverage in measurements. In addition to the
best-fitting, shorter-period orbit, in Fig~\ref{fig:MPcontour} we
provide a $\chi^2$ contour plot showing the correlation between $P$ and
\msini, similar to Figure 3 of
\citet{wright09}. The gray-scale shows the minimum value of $\chi^2$
for single-planet Keplerian fits at fixed values of period and minimum
planet mass. The solid contours denote locations at which $\chi^2$
increases by factors of $\{1,4,9\}$ from inside-out. The dashed
contours show constant 
  eccentricities $e = \{0.2,0.6,0.9\}$ from left to right. For periods
  $P < 100$~years 
  the $\approx 99$\% upper limit on \msini\ is 20~\mjup, with an
  extremely high eccentricity near $e = 0.9$. For eccentricities $e <
  0.6$, \msini~$ < 13$~\mjup\ at roughly 68\% confidence and
  \msini~$ < 15$~\mjup at $\approx$99\% confidence. Given the rarity
  of known planets with 
  \msini~$ > 10$~\mjup\ around stars with masses $M_\star < 2$~\msun,
  it is likely that the true mass of \starP\,b is near or below the
  dueterium-burning limit \citep{spiegel11}. 

{\it \starR, \starB, \starE}---These stars exhibit RV scatter well in
excess of the mean jitter value of 5~\ms\ reported by
\citet{johnson10a}. In all cases the excess scatter may be due to
additional orbital companions. However, periodograms of the residuals
about the best-fitting Keplerian models reveal no convincing additional
periodicities. Examination of the residuals of \starB\ shows that the
tallest periodogram peaks are near 
30~days, and 50~days, with both periodicities below the 1\% 
false-alarm probability (FAP) level. For the residuals of
\starE\ there is a strong peak near $P=900$~days with
FAP~$=0.8$\%. \starR\ similarly shows a strong peak near 800~days
with FAP~$ \sim 1$\%. Additional  
monitoring is warranted for these systems, 
as well as those with linear RV trends.

\section{Summary and Discussion}
\label{summary_and_discussion}

We have reported precise Doppler-shift measurements of
\npllet\ subgiant stars, revealing evidence of Jovian-mass planetary
companions. The host stars of these planets span a wide range of
masses and chemical composition, and thereby provide additional
leverage for studying the relationships between the physical
characteristics of stars and their planets. Evolved intermediate-mass
stars ($M_\star > 1.5$~\msun) have 
proven to be particularly valuable in this regard, providing a much
needed extension of exoplanet discovery space to higher stellar masses
than can be studied on the main sequence, while simultaneously
providing a remarkably large windfall of giant planets. 

The \npllet\ new planets announced herein further highlight the
differences between the known population of planets around evolved,
intermediate-mass stars and those found orbiting Sun-like stars. The
initial discoveries of planets around retired A-type stars revealed a
marked decreased occrrence of planets inward of 1~AU. Indeed, there
are no planets known to orbit between 0.1~AU and 0.6~AU around stars
with $M_\star > 1.5$~\msun. 

The large number of detections from our sample are a testament to
the planet-enriched environs around stars more massive than the
Sun. \citet{johnson10c} used the preliminary detections of the planets
announced in this contribution, along with the
detections from the CPS Doppler surveys of less massive dwarf stars,
to measure the rate of planet occurrence versus stellar mass and
metallicity. They found that at fixed metallicity, the number of stars
harboring a gas giant planet (\msini~$ \gtrsim 0.5$~\mjup) with $a <
3$~AU rises approximately 
linearly with stellar mass. And just as had been measured previously
for Sun-like stars \citep{gonzales97,santos04,fischer05b}, Johnson et
al. found evidence of a planet-metallicity correlation among their
more diverse sample of stars. 

These observed correlations between stellar properties and giant
planet occurrence provide strong constraints for
theories of planet formation. Any successful formation mechanism must
not only describe the formation of the planets in our Solar System,
but must also account for the ways in which planet occurrence varies
with stellar mass and chemical composition. The link between planet
occurrence and stellar properties may be related to the relationship
between stars and their natal circumstellar disks. More massive,
metal-rich stars likely had more massive, dust-enriched protoplanetary
disks that more efficiently form embryonic solid cores that in turn
sweep up gas, resulting in the gas giants detected today. 

The correlation between stellar mass and exoplanets also points the
way toward future discoveries using techniques that are complementary
to Doppler detection. To identify the best targets for high-contrast
imaging surveys, \citet{crepp11} extrapolated to larger semimajor axes
the occurrence
rates and other correlations between stellar and planetary properties
from Doppler surveys. Based on their
Monte Carlo simulations of nearby stars
Crepp \& Johnson found that A type stars are likely to be 
promising targets for the next generation imaging surveys such as the
\emph{Gemini 
Planet Imager}, \emph{Sphere} and {\emph Project 1640}
\citep{gpi,sphere,hinkley11}. According to 
their simulations, the relative discovery rate of planets around A
stars versus M stars will, in relatively short order, help discern the
mode of formation for planets in wide ($a \gtrsim 10$~AU) orbits. For
example, an overabundance of massive planets in wide orbits around A
stars as compared to discoveries around M dwarfs will indicate that
the same formation mechanism responsible for the Doppler-detected
sample of gas giants operates at much wider separations. Thus, just as
the first handful of planets discovered by Doppler surveys revealed
the planet-metallicity relationship now familiar today, the first
handful of directly imaged planets will provide valuable insight into
the stellar mass dependence of the formation of widely orbiting
planets. 

Additional planets from all types of planet-search programs will
enlarge sample sizes and reveal additional, telling correlations and
peculiarities. As the time baselines of Doppler surveys increase,
planets at ever wider semimajor axes will be discovered, revealing the
populations of planets that have not moved far from their birth
places. As Doppler surveys move outward they will be complemented by
increases in the sensitives of direct imaging surveys searching for
planets closer to their host stars and at lower and lower masses. This
overlap will most likely happen the quickest around A stars, both
main-sequence and retired, providing valuable information about planet
formation over four orders of magnitude in semimajor axis.  

\begin{deluxetable}{lll}
\tablecaption{Radial Velocities for HD 5891\label{vel5891}}
\tablewidth{0pt}
\tablehead{
\colhead{HJD} &
\colhead{RV} &
\colhead{Uncertainty} \\
\colhead{-2440000} &
\colhead{(m~s$^{-1}$)} &
\colhead{(m~s$^{-1}$)} 
}
\startdata
14339.9257 &    0.00 &  1.32 \\
14399.8741 &   14.40 &  1.41 \\
14675.0029 & -101.80 &  1.33 \\
14717.9867 &   59.72 &  1.40 \\
15015.0521 & -229.51 &  1.32 \\
15017.1151 & -252.49 &  1.36 \\
15048.9850 &  -47.12 &  1.40 \\
15075.1005 &   77.03 &  1.48 \\
15076.0885 &   84.20 &  1.23 \\
15077.0766 &   88.74 &  1.37 \\
15078.0802 &   89.43 &  1.35 \\
15079.0828 &  104.30 &  1.40 \\
15111.8798 &  -40.60 &  1.51 \\
15135.0589 & -113.11 &  1.30 \\
15171.9551 & -234.11 &  1.63 \\
15187.7896 & -246.19 &  1.40 \\
15196.7607 & -148.30 &  1.35 \\
15198.8379 & -184.96 &  1.41 \\
15229.7277 &   13.78 &  1.29 \\
15250.7176 &  124.93 &  1.45 \\
15255.7218 &   65.51 &  1.40 \\
15396.1245 &  -91.80 &  1.43 \\
15404.1162 &  -66.27 &  1.26 \\
15405.0745 &  -20.21 &  1.39 \\
15406.0816 &  -41.20 &  1.26 \\
15407.1010 &  -31.38 &  1.21 \\
15412.0014 &   26.94 &  1.22 \\
15414.0290 &   61.79 &  1.20 \\
15426.1218 &  103.97 &  1.22 \\
15426.9953 &  112.44 &  1.23 \\
15427.9420 &  118.27 &  1.37 \\
15429.0078 &  115.51 &  1.05 \\
15432.1135 &   95.23 &  1.23 \\
15433.0935 &   78.87 &  1.18 \\
15433.9790 &   61.60 &  1.19 \\
15435.0523 &   69.27 &  1.32 \\
15438.0958 &   10.98 &  1.27 \\
15439.0659 &   64.64 &  1.37 \\
15455.9425 &   25.71 &  1.37 \\
15467.1200 &    6.21 &  1.45 \\
15469.1005 &  -43.24 &  1.45 \\
15470.0312 &  -67.08 &  1.24 \\
15471.7947 &  -75.53 &  1.36 \\
15487.0751 & -165.74 &  1.39 \\
15489.9608 & -183.77 &  1.56 \\
15490.7952 & -198.38 &  1.37 \\
15500.8587 & -286.77 &  1.40 \\
15522.8948 & -274.58 &  1.39 \\
15528.8646 & -234.13 &  1.34 \\
15542.8572 & -255.45 &  1.30 \\
15544.9247 & -261.59 &  1.56 \\
15584.7351 &    5.02 &  1.49 \\
15613.7133 &  140.41 &  2.40 \\
15731.1086 & -183.76 &  1.24
\\
\enddata
\end{deluxetable}

\begin{deluxetable}{llll}
\tablecaption{Radial Velocities for HD 1502\label{vel1502}}
\tablewidth{0pt}
\tablehead{
\colhead{JD} &
\colhead{RV} &
\colhead{Uncertainty} &
\colhead{Telescope} \\
\colhead{-2440000} &
\colhead{(m~s$^{-1}$)} &
\colhead{(m~s$^{-1}$)} &
\colhead{}
}
\startdata
14339.927 &  -40.99 &  1.81 & K \\
14399.840 &  -37.37 &  1.92 & K \\
14455.835 &   11.74 &  1.74 & K \\
14675.004 &   40.95 &  1.82 & K \\
14689.001 &   43.01 &  1.82 & K \\
14717.944 &   20.39 &  1.85 & K \\
14722.893 &   11.99 &  1.85 & K \\
14777.880 &  -15.02 &  1.78 & K \\
14781.811 &  -32.99 &  3.21 & M \\
14790.879 &  -19.32 &  1.68 & K \\
14805.805 &  -21.62 &  1.71 & K \\
14838.766 &   -3.59 &  1.65 & K \\
14841.588 &  -15.98 &  2.83 & M \\
14846.742 &   -8.77 &  1.84 & K \\
14866.725 &   -6.17 &  3.21 & K \\
14867.739 &   -3.89 &  1.97 & K \\
14987.119 &   65.62 &  1.74 & K \\
15015.051 &   65.31 &  1.84 & K \\
15016.082 &   65.49 &  1.70 & K \\
15019.057 &   68.57 &  1.96 & K \\
15024.909 &   77.15 &  2.40 & M \\
15027.910 &   67.88 &  4.04 & M \\
15029.082 &   80.82 &  1.89 & K \\
15045.070 &   83.04 &  1.76 & K \\
15053.900 &   54.41 &  4.42 & M \\
15072.887 &   40.18 &  3.92 & M \\
15076.088 &   58.50 &  1.69 & K \\
15081.093 &   54.00 &  1.73 & K \\
15084.145 &   46.98 &  1.87 & K \\
15109.892 &   45.93 &  2.02 & K \\
15133.977 &   16.31 &  1.76 & K \\
15135.771 &   12.49 &  1.54 & K \\
15135.819 &  -30.75 &  6.17 & M \\
15152.711 &  -13.01 &  2.33 & M \\
15154.769 &    5.06 &  1.56 & M \\
15169.858 &  -14.73 &  1.76 & K \\
15171.883 &  -12.61 &  1.69 & K \\
15172.717 &  -30.42 &  4.45 & M \\
15172.844 &  -18.63 &  1.70 & K \\
15177.688 &  -51.52 &  5.61 & H \\
15181.669 &  -58.04 &  5.12 & H \\
15182.661 &  -64.16 &  3.60 & H \\
15183.657 &  -60.87 &  2.33 & H \\
15185.651 &  -58.92 &  3.68 & H \\
15187.851 &  -32.87 &  1.78 & K \\
15188.646 &  -64.54 &  3.57 & H \\
15188.889 &  -31.96 &  1.74 & K \\
15189.779 &  -29.21 &  1.59 & K \\
15190.646 &  -67.86 &  2.61 & H \\
15190.775 &  -31.88 &  1.69 & K \\
15193.642 &  -65.59 &  3.27 & H \\
15196.758 &  -33.47 &  1.57 & K \\
15197.784 &  -27.93 &  1.78 & K \\
15198.801 &  -25.23 &  1.56 & K \\
15202.598 &  -68.19 &  1.93 & H \\
15209.584 &  -68.01 &  3.66 & H \\
15221.599 &  -51.96 &  3.68 & M \\
15222.588 &  -57.10 &  2.79 & M \\
15223.577 &  -57.76 &  3.76 & M \\
15226.568 &  -58.24 &  3.83 & M \\
15227.568 &  -49.93 &  2.95 & M \\
15229.725 &  -45.43 &  1.59 & K \\
15231.745 &  -43.78 &  1.74 & K \\
15250.715 &  -40.69 &  1.65 & K \\
15256.709 &  -33.45 &  1.90 & K \\
15377.122 &   70.88 &  1.84 & K \\
15405.078 &   82.90 &  1.85 & K \\
15432.807 &   64.92 &  2.94 & H \\
15435.106 &   98.34 &  1.83 & K \\
15436.902 &   78.92 &  3.02 & M \\
15439.994 &   92.33 &  2.00 & K \\
15455.963 &   84.41 &  1.70 & K \\
15468.886 &   59.36 &  4.34 & H \\
15468.888 &   65.31 &  4.23 & M \\
15487.058 &   66.84 &  1.80 & K \\
15491.803 &   24.25 &  3.66 & H \\
15493.853 &   68.22 &  5.03 & M \\
15497.724 &   68.26 &  3.30 & M \\
15501.713 &   42.59 &  3.69 & M \\
15506.780 &   34.25 &  2.46 & H \\
15511.764 &   22.16 &  3.27 & H \\
15519.727 &   21.96 &  2.76 & H \\
15521.871 &   50.61 &  1.79 & K \\
15527.793 &   62.95 &  4.40 & M \\
15528.705 &   65.33 &  4.01 & M \\
15531.704 &   23.08 &  5.55 & H \\
15543.686 &    7.39 &  5.70 & H \\
15547.698 &   24.28 &  4.99 & M \\
15550.578 &   32.74 &  5.69 & M \\
15554.632 &    0.00 &  2.62 & H \\
15565.601 &  -11.84 &  2.49 & H \\
15584.578 &    0.00 &  2.54 & M \\
15584.732 &    9.77 &  1.65 & K \\
15613.708 &   -6.51 &  1.89 & K
\\
\enddata
\end{deluxetable}

\begin{deluxetable}{lll}
\tablecaption{Radial Velocities for HD 18742\label{vel18742}}
\tablewidth{0pt}
\tablehead{
\colhead{HJD} &
\colhead{RV} &
\colhead{Uncertainty} \\
\colhead{-2440000} &
\colhead{(m~s$^{-1}$)} &
\colhead{(m~s$^{-1}$)} 
}
\startdata
14340.1033 &   12.26 &  1.67 \\
14399.9466 &    0.00 &  1.45 \\
14458.8263 &  -46.91 &  1.39 \\
14690.0728 &  -43.10 &  1.63 \\
14719.1392 &  -19.07 &  1.36 \\
14780.0038 &   -4.84 &  1.69 \\
14790.9573 &   -4.31 &  1.60 \\
14805.9154 &   10.55 &  1.63 \\
14838.8094 &   22.88 &  1.55 \\
14846.8698 &   35.60 &  1.60 \\
15077.0932 &   30.15 &  1.54 \\
15109.9772 &   24.72 &  1.79 \\
15134.9911 &   15.75 &  1.64 \\
15171.9048 &   -7.05 &  1.65 \\
15187.8918 &  -12.06 &  1.54 \\
15229.7703 &  -19.43 &  1.38 \\
15255.7374 &  -30.72 &  1.34 \\
15406.1179 &  -35.69 &  1.54 \\
15437.1255 &  -18.68 &  1.47 \\
15465.0699 &  -12.62 &  1.48 \\
15487.0861 &   -1.40 &  1.54 \\
15521.8936 &    1.16 &  1.47 \\
15545.8251 &   10.13 &  1.51 \\
15555.8882 &    0.54 &  1.45 \\
15584.8925 &   12.24 &  1.44 \\
15614.7626 &   10.35 &  1.69
\\
\enddata
\end{deluxetable}

\begin{deluxetable}{lll}
\tablecaption{Radial Velocities for HD 28678\label{vel28678}}
\tablewidth{0pt}
\tablehead{
\colhead{HJD} &
\colhead{RV} &
\colhead{Uncertainty} \\
\colhead{-2440000} &
\colhead{(m~s$^{-1}$)} &
\colhead{(m~s$^{-1}$)} 
}
\startdata
14340.0851 &  -29.22 &  1.40 \\
14399.9815 &  -59.79 &  1.47 \\
14718.1237 &  -32.53 &  1.46 \\
14846.9513 &  -48.79 &  1.72 \\
15080.1261 &   -7.59 &  1.35 \\
15109.9872 &  -34.93 &  1.54 \\
15134.0167 &  -36.54 &  1.60 \\
15171.9170 &  -50.92 &  1.58 \\
15190.9001 &  -55.70 &  1.71 \\
15231.9479 &  -32.09 &  1.64 \\
15260.7954 &  -32.63 &  1.57 \\
15312.7181 &    2.99 &  1.75 \\
15411.1309 &    0.00 &  1.43 \\
15412.1283 &    9.68 &  1.39 \\
15413.1357 &    2.29 &  1.45 \\
15414.1303 &   12.13 &  1.63 \\
15415.1355 &    0.68 &  1.30 \\
15426.1389 &    9.15 &  1.33 \\
15427.1383 &   10.25 &  1.25 \\
15429.1145 &    7.58 &  1.39 \\
15432.1434 &   22.34 &  1.35 \\
15433.1437 &   10.10 &  1.36 \\
15434.1407 &    8.09 &  1.19 \\
15436.1283 &    8.63 &  1.42 \\
15437.1348 &    8.44 &  1.18 \\
15456.0133 &    1.26 &  1.29 \\
15521.8977 &  -45.74 &  1.43 \\
15546.0604 &  -54.63 &  1.58 \\
15584.7710 &  -51.43 &  1.46 \\
15633.8041 &  -28.13 &  1.64
\\
\enddata
\end{deluxetable}

\begin{deluxetable}{lll}
\tablecaption{Radial Velocities for HD 30856\label{vel30856}}
\tablewidth{0pt}
\tablehead{
\colhead{HJD} &
\colhead{RV} &
\colhead{Uncertainty} \\
\colhead{-2440000} &
\colhead{(m~s$^{-1}$)} &
\colhead{(m~s$^{-1}$)} 
}
\startdata
14340.1164 &  -14.83 &  1.51 \\
14399.9763 &  -24.80 &  1.32 \\
14461.8650 &    0.00 &  1.52 \\
14846.9638 &   41.34 &  1.44 \\
15080.1345 &   -2.25 &  1.35 \\
15135.1077 &  -14.83 &  1.30 \\
15172.9305 &  -21.25 &  1.51 \\
15196.8024 &   -9.52 &  1.38 \\
15231.8167 &  -27.79 &  1.27 \\
15255.7427 &  -28.34 &  1.38 \\
15436.1263 &   12.10 &  1.40 \\
15469.1301 &   15.96 &  1.47 \\
15489.9870 &   23.74 &  1.56 \\
15522.9321 &   29.95 &  1.40 \\
15584.9116 &   30.94 &  1.40 \\
15633.8089 &   38.45 &  1.43
\\
\enddata
\end{deluxetable}

\begin{deluxetable}{lll}
\tablecaption{Radial Velocities for HD 33142\label{vel33142}}
\tablewidth{0pt}
\tablehead{
\colhead{HJD} &
\colhead{RV} &
\colhead{Uncertainty} \\
\colhead{-2440000} &
\colhead{(m~s$^{-1}$)} &
\colhead{(m~s$^{-1}$)} 
}
\startdata
14340.1181 &  -37.08 &  1.38 \\
14400.0322 &  -54.93 &  1.31 \\
14461.8774 &  -23.40 &  1.51 \\
14718.1494 &  -40.56 &  1.34 \\
14791.0832 &   -5.15 &  1.19 \\
14806.9547 &   17.49 &  2.78 \\
14839.0184 &   12.02 &  1.44 \\
14846.9625 &    9.60 &  1.55 \\
14864.9169 &   28.72 &  1.45 \\
14929.7214 &    8.43 &  1.32 \\
15076.1199 &  -23.55 &  1.27 \\
15085.0877 &  -26.62 &  1.29 \\
15110.1324 &  -20.92 &  1.87 \\
15173.0550 &   15.53 &  1.48 \\
15187.9049 &   10.43 &  1.34 \\
15188.9637 &   10.84 &  1.41 \\
15189.8282 &    0.59 &  1.33 \\
15190.9016 &    7.61 &  1.51 \\
15196.8131 &    7.81 &  1.29 \\
15197.9728 &    8.34 &  1.36 \\
15199.0003 &    0.00 &  1.44 \\
15229.7751 &    3.44 &  1.50 \\
15255.7479 &   -6.24 &  1.24 \\
15285.7787 &  -29.24 &  1.53 \\
15312.7215 &  -40.37 &  1.33 \\
15429.1118 &  -27.00 &  1.47 \\
15456.0439 &  -13.63 &  1.30 \\
15490.9600 &   -2.03 &  1.36 \\
15521.9705 &    1.97 &  1.35 \\
15546.0736 &   14.34 &  1.35 \\
15556.0750 &   14.02 &  1.32 \\
15584.9149 &   -0.43 &  1.33 \\
15633.8113 &  -33.00 &  1.35
\\
\enddata
\end{deluxetable}

\begin{deluxetable}{lll}
\tablecaption{Radial Velocities for HD 82886\label{vel82886}}
\tablewidth{0pt}
\tablehead{
\colhead{HJD} &
\colhead{RV} &
\colhead{Uncertainty} \\
\colhead{-2440000} &
\colhead{(m~s$^{-1}$)} &
\colhead{(m~s$^{-1}$)} 
}
\startdata
14216.7876 &  -41.63 &  1.22 \\
14400.1235 &    0.00 &  2.34 \\
14428.1124 &   17.44 &  1.45 \\
14791.1050 &  -19.20 &  2.31 \\
14847.1152 &  -40.40 &  1.48 \\
14865.0033 &  -42.27 &  1.29 \\
14963.8627 &  -23.40 &  1.23 \\
14983.7592 &  -14.35 &  1.27 \\
14984.8081 &  -11.93 &  1.45 \\
14985.8027 &  -22.37 &  1.49 \\
14987.7450 &  -18.10 &  1.29 \\
15014.7406 &  -11.53 &  1.50 \\
15112.1393 &   12.95 &  1.74 \\
15134.1118 &   12.25 &  2.36 \\
15164.1292 &   29.02 &  1.27 \\
15172.1315 &   31.05 &  1.95 \\
15173.1642 &   24.36 &  2.12 \\
15196.9713 &   37.60 &  1.35 \\
15229.0787 &   28.24 &  1.31 \\
15255.7583 &   21.31 &  1.35 \\
15284.8623 &   31.97 &  1.51 \\
15312.8441 &   30.42 &  1.28 \\
15342.7634 &   20.34 &  1.24 \\
15372.7433 &  -11.94 &  1.55 \\
15522.1045 &  -15.76 &  2.44 \\
15556.1032 &   -2.70 &  1.41 \\
15606.9744 &  -23.60 &  1.40 \\
15723.7522 &   13.02 &  1.51
\\
\enddata
\end{deluxetable}

\begin{deluxetable}{lll}
\tablecaption{Radial Velocities for HD 96063\label{vel96063}}
\tablewidth{0pt}
\tablehead{
\colhead{HJD} &
\colhead{RV} &
\colhead{Uncertainty} \\
\colhead{-2440000} &
\colhead{(m~s$^{-1}$)} &
\colhead{(m~s$^{-1}$)} 
}
\startdata
14216.8469 &  -13.32 &  1.45 \\
14544.0373 &   11.18 &  1.49 \\
14847.0517 &    1.92 &  1.61 \\
14988.8574 &  -21.00 &  1.44 \\
15172.1472 &   -8.24 &  2.61 \\
15189.1169 &    0.40 &  2.73 \\
15232.1399 &   11.30 &  1.56 \\
15261.0068 &    8.76 &  1.47 \\
15285.8725 &    6.51 &  1.34 \\
15320.7818 &  -21.58 &  1.42 \\
15344.7905 &  -32.46 &  1.50 \\
15372.7605 &  -34.90 &  1.43 \\
15403.7335 &  -46.03 &  1.43 \\
15605.9963 &    0.00 &  1.46 \\
15615.0563 &    1.31 &  1.57
\\
\enddata
\end{deluxetable}

\begin{deluxetable}{lll}
\tablecaption{Radial Velocities for HD 98219\label{vel98219}}
\tablewidth{0pt}
\tablehead{
\colhead{HJD} &
\colhead{RV} &
\colhead{Uncertainty} \\
\colhead{-2440000} &
\colhead{(m~s$^{-1}$)} &
\colhead{(m~s$^{-1}$)} 
}
\startdata
14216.8449 &   57.96 &  1.13 \\
14544.0427 &   19.96 &  1.52 \\
14640.7494 &   51.97 &  2.09 \\
14847.0563 &  -32.03 &  1.41 \\
14983.7787 &   19.55 &  1.29 \\
15171.1635 &   10.20 &  1.33 \\
15189.1341 &    4.18 &  2.48 \\
15229.0611 &  -16.08 &  1.30 \\
15252.0420 &  -23.02 &  1.26 \\
15255.8857 &  -23.18 &  1.24 \\
15285.8704 &  -19.25 &  1.30 \\
15313.8334 &  -25.03 &  1.29 \\
15342.7941 &  -12.41 &  1.26 \\
15376.7390 &    0.00 &  1.25 \\
15585.1462 &   33.03 &  1.33 \\
15606.0328 &   16.94 &  1.28 \\
15700.7708 &  -33.64 &  1.23
\\
\enddata
\end{deluxetable}

\begin{deluxetable}{lll}
\tablecaption{Radial Velocities for HD 99706\label{vel99706}}
\tablewidth{0pt}
\tablehead{
\colhead{HJD} &
\colhead{RV} &
\colhead{Uncertainty} \\
\colhead{-2440000} &
\colhead{(m~s$^{-1}$)} &
\colhead{(m~s$^{-1}$)} 
}
\startdata
14428.1613 &   38.42 &  1.27 \\
14429.0887 &   35.89 &  1.29 \\
14464.0610 &   26.33 &  1.49 \\
14847.0764 &    4.66 &  1.39 \\
14988.8390 &    0.00 &  1.29 \\
15174.1629 &   33.84 &  1.23 \\
15229.0719 &   34.84 &  1.22 \\
15255.9592 &   31.65 &  1.18 \\
15256.9781 &   28.77 &  1.20 \\
15284.9159 &   23.61 &  1.28 \\
15313.9492 &   16.02 &  1.19 \\
15343.8495 &   12.69 &  1.21 \\
15378.7470 &   -2.49 &  1.30 \\
15404.7331 &   -0.41 &  1.23 \\
15543.1727 &   -8.84 &  1.19 \\
15585.1087 &  -12.15 &  1.15 \\
15605.9817 &  -25.72 &  1.33 \\
15615.0400 &  -20.42 &  1.35 \\
15633.9015 &  -20.59 &  1.27 \\
15663.9636 &  -12.73 &  1.20 \\
15667.9704 &  -12.98 &  1.18 \\
15700.8151 &  -19.50 &  1.20 \\
15734.7675 &  -15.18 &  1.11 \\
15770.7434 &  -13.48 &  1.37
\\
\enddata
\end{deluxetable}

\begin{deluxetable}{lll}
\tablecaption{Radial Velocities for HD 102329\label{vel102329}}
\tablewidth{0pt}
\tablehead{
\colhead{HJD} &
\colhead{RV} &
\colhead{Uncertainty} \\
\colhead{-2440000} &
\colhead{(m~s$^{-1}$)} &
\colhead{(m~s$^{-1}$)} 
}
\startdata
14216.8413 &  -60.44 &  1.20 \\
14544.0391 &   19.90 &  1.23 \\
14847.0691 &   28.39 &  1.32 \\
14988.8623 &  -68.46 &  1.26 \\
15015.8197 &  -84.27 &  1.15 \\
15044.7343 & -107.42 &  1.19 \\
15172.1434 &  -78.90 &  2.51 \\
15189.1143 &  -72.33 &  2.31 \\
15229.0638 &  -51.88 &  1.22 \\
15255.8881 &  -28.90 &  1.15 \\
15289.9324 &   10.79 &  1.18 \\
15313.7792 &   10.12 &  1.14 \\
15342.7984 &   21.54 &  1.10 \\
15373.7400 &   38.91 &  1.22 \\
15402.7556 &   31.52 &  1.30 \\
15606.0374 &   29.79 &  1.15 \\
15615.0521 &   21.05 &  1.29 \\
15633.8900 &    0.00 &  1.09 \\
15667.9885 &  -11.60 &  1.18 \\
15700.7689 &   -8.16 &  1.26
\\
\enddata
\end{deluxetable}

\begin{deluxetable}{lll}
\tablecaption{Radial Velocities for HD 106270\label{vel106270}}
\tablewidth{0pt}
\tablehead{
\colhead{HJD} &
\colhead{RV} &
\colhead{Uncertainty} \\
\colhead{-2440000} &
\colhead{(m~s$^{-1}$)} &
\colhead{(m~s$^{-1}$)} 
}
\startdata
14216.8355 &   23.22 &  1.38 \\
14455.1699 &  102.95 &  1.60 \\
14635.7675 &  171.13 &  1.59 \\
14927.9701 &  164.37 &  1.75 \\
15015.8167 &  121.36 &  1.39 \\
15044.7394 &   81.25 &  1.67 \\
15173.1299 &   34.41 &  2.60 \\
15189.1367 &   28.32 &  2.63 \\
15261.0094 &    0.00 &  1.59 \\
15289.9478 &    8.14 &  1.56 \\
15313.8402 &   -8.69 &  1.59 \\
15342.8011 &  -21.05 &  1.56 \\
15372.7641 &  -39.39 &  1.52 \\
15403.7637 &  -34.73 &  1.62 \\
15585.1385 &  -38.04 &  2.66 \\
15606.0379 &  -70.38 &  1.61 \\
15607.0551 &  -68.71 &  1.59 \\
15633.8911 &  -58.47 &  1.57 \\
15663.8909 &  -64.68 &  1.65 \\
15700.7732 &  -70.54 &  1.41
\\
\enddata
\end{deluxetable}

\begin{deluxetable}{lll}
\tablecaption{Radial Velocities for HD 108863\label{vel108863}}
\tablewidth{0pt}
\tablehead{
\colhead{HJD} &
\colhead{RV} &
\colhead{Uncertainty} \\
\colhead{-2440000} &
\colhead{(m~s$^{-1}$)} &
\colhead{(m~s$^{-1}$)} 
}
\startdata
14216.8115 &  -51.43 &  1.12 \\
14544.0479 &    0.00 &  1.41 \\
14635.8266 &  -46.44 &  1.00 \\
14934.8653 &   31.90 &  1.47 \\
14963.9885 &   24.25 &  1.31 \\
14983.8915 &   13.21 &  1.21 \\
15014.7747 &  -13.40 &  1.19 \\
15016.8627 &  -17.30 &  1.21 \\
15043.7450 &  -30.01 &  1.51 \\
15172.1413 &  -28.18 &  1.52 \\
15189.1458 &  -21.38 &  2.95 \\
15255.8905 &   20.25 &  1.21 \\
15284.8821 &   41.28 &  1.42 \\
15311.8035 &   39.04 &  1.50 \\
15342.8757 &   40.04 &  1.24 \\
15376.7831 &   32.40 &  1.14 \\
15402.7515 &   13.40 &  1.26 \\
15522.1564 &  -42.12 &  2.78 \\
15546.1659 &  -56.68 &  1.29 \\
15585.0903 &  -46.64 &  1.25 \\
15605.9932 &  -21.93 &  1.33 \\
15634.0020 &  -16.91 &  1.30 \\
15663.9420 &    4.31 &  1.30 \\
15704.8126 &   11.83 &  1.25
\\
\enddata
\end{deluxetable}

\begin{deluxetable}{lll}
\tablecaption{Radial Velocities for HD 116029\label{vel116029}}
\tablewidth{0pt}
\tablehead{
\colhead{HJD} &
\colhead{RV} &
\colhead{Uncertainty} \\
\colhead{-2440000} &
\colhead{(m~s$^{-1}$)} &
\colhead{(m~s$^{-1}$)} 
}
\startdata
14216.8196 &  -26.73 &  1.09 \\
14216.9475 &  -24.06 &  0.96 \\
14345.7641 &   20.59 &  1.33 \\
14635.8322 &   -9.41 &  1.02 \\
14954.9950 &   -6.09 &  1.27 \\
14983.9004 &   -2.38 &  1.18 \\
15043.7491 &   29.89 &  1.31 \\
15197.1540 &   44.61 &  1.21 \\
15232.0265 &   21.80 &  1.29 \\
15261.0304 &   28.04 &  1.21 \\
15285.1540 &   20.95 &  1.28 \\
15313.7736 &    4.96 &  1.05 \\
15342.8932 &   -0.75 &  1.18 \\
15379.8156 &  -15.82 &  1.08 \\
15404.7751 &  -25.43 &  1.10 \\
15585.1319 &   -0.25 &  1.13 \\
15605.9881 &   -5.74 &  1.19 \\
15615.0443 &    0.66 &  1.31 \\
15633.9001 &    0.00 &  1.15 \\
15667.9740 &   14.75 &  1.18 \\
15703.7976 &   16.39 &  1.09
\\
\enddata
\end{deluxetable}

\begin{deluxetable}{lll}
\tablecaption{Radial Velocities for HD 131496\label{vel131496}}
\tablewidth{0pt}
\tablehead{
\colhead{HJD} &
\colhead{RV} &
\colhead{Uncertainty} \\
\colhead{-2440000} &
\colhead{(m~s$^{-1}$)} &
\colhead{(m~s$^{-1}$)} 
}
\startdata
14257.7864 &   48.99 &  1.08 \\
14339.7383 &   22.17 &  1.15 \\
14633.8320 &  -11.45 &  1.06 \\
14674.7923 &  -24.32 &  1.19 \\
14964.0661 &   24.79 &  1.08 \\
15041.8436 &   56.78 &  1.30 \\
15042.8783 &   41.12 &  1.28 \\
15081.7136 &   49.43 &  1.19 \\
15197.1629 &   32.75 &  1.29 \\
15231.1524 &   29.08 &  1.29 \\
15257.0130 &   14.66 &  1.35 \\
15284.8831 &    2.07 &  1.35 \\
15314.8533 &    0.83 &  1.46 \\
15343.7940 &    5.12 &  1.22 \\
15379.8222 &  -13.69 &  1.22 \\
15404.7802 &  -19.04 &  1.18 \\
15455.7491 &  -27.49 &  1.24 \\
15546.1613 &  -14.23 &  1.30 \\
15559.1657 &   -2.79 &  1.42 \\
15606.0439 &  -26.97 &  1.19 \\
15607.0571 &  -22.07 &  1.21 \\
15608.0288 &   -8.97 &  1.13 \\
15614.0246 &   -9.23 &  1.16 \\
15615.0451 &  -10.64 &  1.41 \\
15634.0616 &   -6.55 &  1.24 \\
15634.9982 &  -18.47 &  1.35 \\
15635.9791 &   -2.28 &  1.01 \\
15636.9679 &  -16.86 &  1.34 \\
15663.9449 &    0.07 &  1.34 \\
15670.9602 &   -9.97 &  0.63 \\
15671.8330 &    2.24 &  1.03 \\
15672.8200 &    0.63 &  1.19 \\
15673.8340 &    2.12 &  1.25 \\
15697.8643 &    7.12 &  1.43 \\
15698.8647 &    4.48 &  1.29 \\
15699.8226 &   -0.13 &  1.21 \\
15700.8064 &    5.34 &  1.27 \\
15703.7722 &   -2.25 &  1.17 \\
15704.7987 &    4.89 &  1.17 \\
15705.8142 &    0.84 &  1.04 \\
15722.9642 &    0.00 &  1.30
\\
\enddata
\end{deluxetable}

\begin{deluxetable}{lll}
\tablecaption{Radial Velocities for HD 142245\label{vel142245}}
\tablewidth{0pt}
\tablehead{
\colhead{HJD} &
\colhead{RV} &
\colhead{Uncertainty} \\
\colhead{-2440000} &
\colhead{(m~s$^{-1}$)} &
\colhead{(m~s$^{-1}$)} 
}
\startdata
14257.7609 &   -1.70 &  1.00 \\
14339.7424 &  -20.45 &  1.14 \\
14399.6984 &  -11.61 &  1.13 \\
14635.8973 &  -31.07 &  1.20 \\
14674.7988 &  -15.51 &  1.17 \\
14986.8193 &   20.63 &  1.13 \\
15015.9357 &   14.23 &  1.06 \\
15231.1494 &   23.37 &  1.32 \\
15257.0130 &   21.99 &  1.09 \\
15286.0067 &   22.97 &  1.17 \\
15319.9358 &   10.28 &  1.11 \\
15351.8205 &   15.55 &  0.98 \\
15379.7723 &   15.15 &  1.05 \\
15464.7090 &    0.00 &  1.03 \\
15486.6972 &    7.93 &  1.18 \\
15608.0546 &  -11.90 &  1.21 \\
15634.0617 &  -15.12 &  1.19 \\
15700.8053 &  -11.29 &  1.23 \\
15722.7901 &  -22.27 &  1.28
\\
\enddata
\end{deluxetable}

\begin{deluxetable}{lll}
\tablecaption{Radial Velocities for HD 152581\label{vel152581}}
\tablewidth{0pt}
\tablehead{
\colhead{HJD} &
\colhead{RV} &
\colhead{Uncertainty} \\
\colhead{-2440000} &
\colhead{(m~s$^{-1}$)} &
\colhead{(m~s$^{-1}$)} 
}
\startdata
14257.7825 &    0.00 &  1.36 \\
14339.7455 &  -16.51 &  1.49 \\
14399.7115 &  -17.18 &  1.67 \\
14674.8143 &   61.42 &  1.40 \\
14963.8491 &   -7.59 &  1.58 \\
14983.7976 &  -17.46 &  1.67 \\
15043.8501 &  -10.75 &  1.52 \\
15111.7059 &  -13.43 &  1.50 \\
15320.0287 &   45.27 &  1.48 \\
15342.8146 &   51.83 &  1.42 \\
15373.7703 &   56.30 &  1.63 \\
15405.8067 &   49.53 &  1.50 \\
15435.7358 &   54.46 &  1.50 \\
15464.7138 &   46.73 &  1.58 \\
15486.7241 &   34.80 &  1.55 \\
15606.1541 &   -4.01 &  1.41 \\
15607.1341 &    4.78 &  1.49 \\
15608.1201 &   15.57 &  1.46 \\
15613.1531 &    8.45 &  1.39 \\
15614.1706 &   -3.19 &  1.44 \\
15636.0731 &   -2.34 &  1.27 \\
15668.0300 &  -14.89 &  1.51 \\
15706.8514 &  -15.33 &  1.31 \\
15735.8586 &  -16.72 &  1.61
\\
\enddata
\end{deluxetable}

\begin{deluxetable}{lll}
\tablecaption{Radial Velocities for HD 158038\label{vel158038}}
\tablewidth{0pt}
\tablehead{
\colhead{HJD} &
\colhead{RV} &
\colhead{Uncertainty} \\
\colhead{-2440000} &
\colhead{(m~s$^{-1}$)} &
\colhead{(m~s$^{-1}$)} 
}
\startdata
14258.0333 &  -32.20 &  1.11 \\
14287.8726 &  -28.05 &  2.34 \\
14345.8015 &   -4.80 &  1.22 \\
14399.7044 &   14.16 &  1.25 \\
14674.8733 &   -1.99 &  1.17 \\
14955.9772 &   42.82 &  1.33 \\
15014.8461 &   41.88 &  1.34 \\
15028.9808 &   35.15 &  1.30 \\
15111.7333 &    7.37 &  1.25 \\
15135.7129 &   -0.97 &  1.12 \\
15286.0608 &  -10.73 &  1.31 \\
15313.9043 &  -21.31 &  1.15 \\
15342.9667 &  -17.68 &  1.17 \\
15378.7929 &   -4.90 &  1.22 \\
15399.9588 &   -1.89 &  1.24 \\
15405.7727 &   -9.12 &  1.12 \\
15431.7305 &    6.04 &  1.21 \\
15469.7065 &   28.63 &  1.24 \\
15585.1729 &   22.56 &  1.16 \\
15606.1764 &   15.81 &  1.08 \\
15636.0867 &    5.26 &  1.12 \\
15668.0043 &  -11.54 &  1.13 \\
15704.8546 &   -9.72 &  1.16 \\
15722.8918 &  -24.06 &  1.23
\\
\enddata
\end{deluxetable}

\clearpage

\acknowledgments
We thank the many observers who contributed to the observations
reported here. We gratefully acknowledge the efforts and dedication
of the Keck Observatory staff, especially Grant Hill, Scott Dahm and
Hien Tran for their support of 
HIRES and Greg Wirth for support of remote observing. We are also
grateful to the time assignment committees of NASA, NOAO, Caltech, and
the University of California for their generous allocations of
observing time. J.\,A.\,J.\ thanks the NSF Astronomy and Astrophysics
Postdoctoral Fellowship program for support in the years leading to
the completion of this work, and acknowledges support form NSF grant
AST-0702821 and the NASA Exoplanets Science Institute
(NExScI). G.\,W.\,M.\ acknowledges NASA grant NNX06AH52G.   
J.T.W. was partially supported by funding from the Center for Exoplanets
and Habitable Worlds.  The Center for Exoplanets and Habitable Worlds
is supported by the Pennsylvania State University, the Eberly College
of Science, and the Pennsylvania Space Grant Consortium.
G.\,W.\,H acknowledges support from NASA, NSF,
Tennessee State University, and the State of Tennessee through its
Centers of Excellence program. 
Finally, the authors wish to extend special thanks to those of
Hawaiian ancestry  on whose sacred mountain of Mauna Kea we are
privileged to be guests.   
Without their generous hospitality, the Keck observations presented herein
would not have been possible.

\bibliography{}

\begin{thebibliography}{87}

\bibitem[{{Bowler} {et~al.}(2010){Bowler}, {Johnson}, {Marcy}, {Henry}, {Peek},  {Fischer}, {Clubb}, {Liu}, {Reffert}, {Schwab}, \& {Lowe}}]{bowler10}
{Bowler}, B.~P., {et~al.} 2010, \apj, 709, 396

\bibitem[{{Brugamyer} {et~al.}(2011){Brugamyer}, {Dodson-Robinson}, {Cochran},  \& {Sneden}}]{brugamyer11}
{Brugamyer}, E., {et~al.} 2011, ArXiv e-prints

\bibitem[{{Butler} {et~al.}(1996){Butler}, {Marcy}, {Williams}, {McCarthy},  {Dosanjh}, \& {Vogt}}]{butler96}
{Butler}, R.~P., {et~al.} 1996, \pasp, 108, 500

\bibitem[{{Charbonneau} {et~al.}(2007){Charbonneau}, {Brown}, {Burrows}, \&  {Laughlin}}]{charbonneau07}
{Charbonneau}, D., {et~al.} 2007, 701

\bibitem[{{Claudi} {et~al.}(2006){Claudi}, {Turatto}, {Antichi}, {Gratton},  {Scuderi}, {Cascone}, {Mesa}, {Desidera}, {Baruffolo}, {Berton}, {Bagnara},  {Giro}, {Bruno}, {Fantinel}, {Beuzit}, {Puget}, \& {Dohlen}}]{sphere}
{Claudi}, R.~U., {et~al.} 2006, SPIE, 6269

\bibitem[{{Crepp} \& {Johnson}(2011)}]{crepp11}
{Crepp}, J.~R. \& {Johnson}, J.~A. 2011, ArXiv:1103.4910

\bibitem[{{da Silva} {et~al.}(2006){da Silva}, {Udry}, {Bouchy}, {Mayor},  {Moutou}, {Pont}, {Queloz}, {Santos}, {S{\'e}gransan}, \&  {Zucker}}]{dasilva06}
{da Silva}, R., {et~al.} 2006,  \aap, 446, 717

\bibitem[{{Eaton} {et~al.}(2003){Eaton}, {Henry}, \& {Fekel}}]{eaton03}
{Eaton}, J.~A., {Henry}, G.~W., \& {Fekel}, F.~C. 2003, The Future of Small  Telescopes In The New Millennium.~Volume II - The Telescopes We Use, 189

\bibitem[{{Endl} {et~al.}(2003){Endl}, {Cochran}, {Tull}, \&  {MacQueen}}]{endl03}
{Endl}, M., {et~al.} 2003, \aj,  126, 3099

\bibitem[{{Endl} {et~al.}(2000){Endl}, {K{\"u}rster}, \& {Els}}]{endl00}
{Endl}, M., {K{\"u}rster}, M., \& {Els}, S. 2000, \aap, 362, 585

\bibitem[{{Fischer} {et~al.}(2005){Fischer}, {Laughlin}, {Butler}, {Marcy},  {Johnson}, {Henry}, {Valenti}, {Vogt}, {Ammons}, {Robinson}, {Spear},  {Strader}, {Driscoll}, {Fuller}, {Johnson}, {Manrao}, {McCarthy},  {Mu{\~n}oz}, {Tah}, {Wright}, {Ida}, {Sato}, {Toyota}, \&  {Minniti}}]{fischer05a}
{Fischer}, D.~A., {et~al.} 2005, \apj, 620, 481

\bibitem[{{Fischer} \& {Valenti}(2005)}]{fischer05b}
{Fischer}, D.~A. \& {Valenti}, J. 2005, \apj, 622, 1102

\bibitem[{{Flower}(1996)}]{flower96}
{Flower}, P.~J. 1996, \apj, 469, 355

\bibitem[{{Ford}(2005)}]{ford05}
{Ford}, E.~B. 2005, \aj, 129, 1706

\bibitem[{{Ford} \& {Gregory}(2007)}]{ford07}
{Ford}, E.~B. \& {Gregory}, P.~C. 2007, 371, 189

\bibitem[{{Galland} {et~al.}(2005){Galland}, {Lagrange}, {Udry}, {Chelli},  {Pepe}, {Queloz}, {Beuzit}, \& {Mayor}}]{galland05}
{Galland}, F., {et~al.} 2005, \aap, 443, 337

\bibitem[{{Ghezzi} {et~al.}(2010){Ghezzi}, {Cunha}, {Schuler}, \&  {Smith}}]{ghezzi10}
{Ghezzi}, L., {et~al.} 2010, \apj, 725,  721

\bibitem[{{Girardi} {et~al.}(2002){Girardi}, {Bertelli}, {Bressan}, {Chiosi},  {Groenewegen}, {Marigo}, {Salasnich}, \& {Weiss}}]{girardi02}
{Girardi}, L., {et~al.} 2002, \aap, 391, 195

\bibitem[{{Gonzalez}(1997a)}]{gonzalez97}
{Gonzalez}, G. 1997a, \mnras, 285, 403

\bibitem[{{Gonzalez}(1997b)}]{gonzales97}
---. 1997b, \mnras, 285, 403

\bibitem[{{Gregory} \& {Fischer}(2010)}]{gregory10}
{Gregory}, P.~C. \& {Fischer}, D.~A. 2010, \mnras, 403, 731

\bibitem[{{Hatzes} {et~al.}(2003){Hatzes}, {Cochran}, {Endl}, {McArthur},  {Paulson}, {Walker}, {Campbell}, \& {Yang}}]{hatzes03}
{Hatzes}, A.~P., {et~al.} 2003, \apj, 599,  1383

\bibitem[{{Hekker} {et~al.}(2006){Hekker}, {Reffert}, {Quirrenbach},  {Mitchell}, {Fischer}, {Marcy}, \& {Butler}}]{hekker06}
{Hekker}, S., {et~al.} 2006, \aap, 454, 943

\bibitem[{{Henry} {et~al.}(1995a){Henry}, {Eaton}, {Hamer}, \&  {Hall}}]{henry95b}
{Henry}, G.~W., {et~al.} 1995a, \apjs, 97, 513

\bibitem[{{Henry} {et~al.}(1995b){Henry}, {Fekel}, \&  {Hall}}]{henry95}
{Henry}, G.~W., {Fekel}, F.~C., \& {Hall}, D.~S. 1995b, \aj, 110,  2926

\bibitem[{{Henry} {et~al.}(2000){Henry}, {Fekel}, {Henry}, \&  {Hall}}]{henry00c}
{Henry}, G.~W., {et~al.} 2000, \apjs,  130, 201

\bibitem[{{Hinkley} {et~al.}(2011){Hinkley}, {Oppenheimer}, {Zimmerman},  {Brenner}, {Parry}, {Crepp}, {Vasisht}, {Ligon}, {King}, {Soummer},  {Sivaramakrishnan}, {Beichman}, {Shao}, {Roberts}, {Bouchez}, {Dekany},  {Pueyo}, {Roberts}, {Lockhart}, {Zhai}, {Shelton}, \& {Burruss}}]{hinkley11}
{Hinkley}, S., {et~al.} 2011, \pasp, 123, 74

\bibitem[{{Howard} {et~al.}(2011a){Howard}, {Johnson}, {Marcy},  {Fischer}, {Wright}, {Henry}, {Isaacson}, {Valenti}, {Anderson}, \&  {Piskunov}}]{howard11b}
{Howard}, A.~W., {et~al.} 2011a, \apj, 726, 73

\bibitem[{{Howard} {et~al.}(2011b){Howard}, {Marcy}, {Bryson},  {Jenkins}, {Rowe}, {Batalha}, {Borucki}, {Koch}, {Dunham}, {Gautier}, {Van  Cleve}, {Cochran}, {Latham}, {Lissauer}, {Torres}, {Brown}, {Gilliland},  {Buchhave}, {Caldwell}, {Christensen-Dalsgaard}, {Ciardi}, {Fressin}, {Haas},  {Howell}, {Kjeldsen}, {Seager}, {Rogers}, {Sasselov}, {Steffen}, {Basri},  {Charbonneau}, {Christiansen}, {Clarke}, {Dupree}, {Fabrycky}, {Fischer},  {Ford}, {Fortney}, {Tarter}, {Girouard}, {Holman}, {Johnson}, {Klaus},  {Machalek}, {Moorhead}, {Morehead}, {Ragozzine}, {Tenenbaum}, {Twicken},  {Quinn}, {Isaacson}, {Shporer}, {Lucas}, {Walkowicz}, {Welsh}, {Boss},  {Devore}, {Gould}, {Smith}, {Morris}, {Prsa}, \& {Morton}}]{howard11}
{Howard}, A.~W., {et~al.} 2011b, ArXiv:1103.2541

\bibitem[{{Ida} \& {Lin}(2004)}]{ida04}
{Ida}, S. \& {Lin}, D.~N.~C. 2004, \apj, 604, 388

\bibitem[{{Isaacson} \& {Fischer}(2010)}]{isaacson10}
{Isaacson}, H. \& {Fischer}, D. 2010, \apj, 725, 875

\bibitem[{{Johnson}(2008)}]{johnson08c}
{Johnson}, J.~A. 2008, in Astronomical Society of the Pacific Conference  Series, Vol. 398, Extreme Solar Systems, ed. {D.~Fischer, F.~A.~Rasio,  S.~E.~Thorsett, \& A.~Wolszczan}, 59--+

\bibitem[{{Johnson} {et~al.}(2010a){Johnson}, {Aller}, {Howard},  \& {Crepp}}]{johnson10c}
{Johnson}, J.~A., {et~al.} 2010a, \pasp, 122, 905

\bibitem[{{Johnson} {et~al.}(2010b){Johnson}, {Bowler}, {Howard},  {Henry}, {Marcy}, {Isaacson}, {Brewer}, {Fischer}, {Morton}, \&  {Crepp}}]{johnson10d}
{Johnson}, J.~A., {et~al.} 2010b, \apjl, 721, L153

\bibitem[{{Johnson} {et~al.}(2007){Johnson}, {Fischer}, {Marcy}, {Wright},  {Driscoll}, {Butler}, {Hekker}, {Reffert}, \& {Vogt}}]{johnson07}
{Johnson}, J.~A., {et~al.} 2007, \apj, 665, 785

\bibitem[{{Johnson} {et~al.}(2010c){Johnson}, {Howard}, {Bowler},  {Henry}, {Marcy}, {Wright}, {Fischer}, \& {Isaacson}}]{johnson10b}
{Johnson}, J.~A., {et~al.} 2010c, \pasp, 122, 701

\bibitem[{{Johnson} {et~al.}(2010d){Johnson}, {Howard}, {Marcy},  {Bowler}, {Henry}, {Fischer}, {Apps}, {Isaacson}, \& {Wright}}]{johnson10a}
{Johnson}, J.~A., {et~al.} 2010d, \pasp, 122, 149

\bibitem[{{Johnson} {et~al.}(2006){Johnson}, {Marcy}, {Fischer}, {Henry},  {Wright}, {Isaacson}, \& {McCarthy}}]{johnson06b}
{Johnson}, J.~A., {et~al.} 2006, \apj, 652, 1724

\bibitem[{{Johnson} {et~al.}(2011){Johnson}, {Payne}, {Howard}, {Clubb},  {Ford}, {Bowler}, {Henry}, {Fischer}, {Marcy}, {Brewer}, {Schwab}, {Reffert},  \& {Lowe}}]{johnson11a}
{Johnson}, J.~A., {et~al.} 2011, \aj, 141, 16

\bibitem[{{Johnson} {et~al.}(2008){Johnson}, {Winn}, {Narita}, {Enya},  {Williams}, {Marcy}, {Sato}, {Ohta}, {Taruya}, {Suto}, {Turner}, {Bakos},  {Butler}, {Vogt}, {Aoki}, {Tamura}, {Yamada}, {Yoshii}, \&  {Hidas}}]{johnson08b}
{Johnson}, J.~A., {et~al.} 2008, \apj, 686, 649

\bibitem[{{Kennedy} \& {Kenyon}(2008)}]{kennedy08}
{Kennedy}, G.~M. \& {Kenyon}, S.~J. 2008, \apj, 673, 502

\bibitem[{Kuha(2004)}]{kuha04}
Kuha, J. 2004, Sociological Methods Research, 33, 188

\bibitem[{{Lagrange} {et~al.}(2010){Lagrange}, {Desort}, \&  {Meunier}}]{lagrange10}
{Lagrange}, A., {Desort}, M., \& {Meunier}, N. 2010, arXiv:1001.1449

\bibitem[{{Laughlin}(2000)}]{laughlin00}
{Laughlin}, G. 2000, \apj, 545, 1064

\bibitem[{{Laughlin} {et~al.}(2004){Laughlin}, {Bodenheimer}, \&  {Adams}}]{laughlin04}
{Laughlin}, G., {Bodenheimer}, P., \& {Adams}, F.~C. 2004, \apjl, 612, L73

\bibitem[{{Liddle}(2004)}]{liddle04}
{Liddle}, A.~R. 2004, \mnras, 351, L49

\bibitem[{{Lloyd}(2011)}]{lloyd11}
{Lloyd}, J.~P. 2011, ArXiv e-prints

\bibitem[{{Lovis} \& {Mayor}(2007)}]{lovis07}
{Lovis}, C. \& {Mayor}, M. 2007, \aap, 472, 657

\bibitem[{{Macintosh} {et~al.}(2008){Macintosh}, {Graham}, {Palmer}, {Doyon},  {Dunn}, {Gavel}, {Larkin}, {Oppenheimer}, {Saddlemyer}, {Sivaramakrishnan},  {Wallace}, {Bauman}, {Erickson}, {Marois}, {Poyneer}, \& {Soummer}}]{gpi}
{Macintosh}, B.~A., {et~al.} 2008, SPIE, 7015

\bibitem[{{Makarov} {et~al.}(2009){Makarov}, {Beichman}, {Catanzarite},  {Fischer}, {Lebreton}, {Malbet}, \& {Shao}}]{makarov09}
{Makarov}, V.~V., {et~al.} 2009, \apjl, 707, L73

\bibitem[{{Marcy} \& {Butler}(1992)}]{marcy92b}
{Marcy}, G.~W. \& {Butler}, R.~P. 1992, \pasp, 104, 270

\bibitem[{{Marois} {et~al.}(2008){Marois}, {Macintosh}, {Barman}, {Zuckerman},  {Song}, {Patience}, {Lafreni{\`e}re}, \& {Doyon}}]{marois08}
{Marois}, C., {et~al.} 2008, Science, 322, 1348

\bibitem[{{Mordasini} {et~al.}(2009){Mordasini}, {Alibert}, {Benz}, \&  {Naef}}]{mordasini09}
{Mordasini}, C., {et~al.} 2009, \aap, 501, 1161

\bibitem[{{Pasquini} {et~al.}(2007){Pasquini}, {D{\"o}llinger}, {Weiss},  {Girardi}, {Chavero}, {Hatzes}, {da Silva}, \& {Setiawan}}]{pasquini07}
{Pasquini}, L., {et~al.} 2007, \aap, 473, 979

\bibitem[{{Paulson} {et~al.}(2004){Paulson}, {Saar}, {Cochran}, \&  {Henry}}]{paulson04}
{Paulson}, D.~B., {et~al.} 2004, \aj,  127, 1644

\bibitem[{{Peek} {et~al.}(2009){Peek}, {Johnson}, {Fischer}, {Marcy}, {Henry},  {Howard}, {Wright}, {Lowe}, {Reffert}, {Schwab}, {Williams}, {Isaacson}, \&  {Giguere}}]{peek09}
{Peek}, K.~M.~G., {et~al.} 2009,  \pasp, 121, 613

\bibitem[{{Queloz} {et~al.}(2001){Queloz}, {Henry}, {Sivan}, {Baliunas},  {Beuzit}, {Donahue}, {Mayor}, {Naef}, {Perrier}, \& {Udry}}]{queloz01b}
{Queloz}, D., {et~al.} 2001, \aap, 379, 279

\bibitem[{{Ramsey} {et~al.}(1998){Ramsey}, {Adams}, {Barnes}, {Booth},  {Cornell}, {Fowler}, {Gaffney}, {Glaspey}, {Good}, {Hill}, {Kelton},  {Krabbendam}, {Long}, {MacQueen}, {Ray}, {Ricklefs}, {Sage}, {Sebring},  {Spiesman}, \& {Steiner}}]{ramsey98}
{Ramsey}, L.~W., {et~al.} 1998, in Society of Photo-Optical  Instrumentation Engineers (SPIE) Conference Series, Vol. 3352, Society of  Photo-Optical Instrumentation Engineers (SPIE) Conference Series, ed.  {L.~M.~Stepp}, 34--42

\bibitem[{{Reffert} {et~al.}(2006){Reffert}, {Quirrenbach}, {Mitchell},  {Albrecht}, {Hekker}, {Fischer}, {Marcy}, \& {Butler}}]{reffert06}
{Reffert}, S., {et~al.} 2006, \apj, 652, 661

\bibitem[{{Saar} {et~al.}(1998){Saar}, {Butler}, \& {Marcy}}]{saar98}
{Saar}, S.~H., {Butler}, R.~P., \& {Marcy}, G.~W. 1998, \apjl, 498, L153+

\bibitem[{{Santos} {et~al.}(2004){Santos}, {Israelian}, \& {Mayor}}]{santos04}
{Santos}, N.~C., {Israelian}, G., \& {Mayor}, M. 2004, \aap, 415, 1153

\bibitem[{{Sato} {et~al.}(2005){Sato}, {Fischer}, {Henry}, {Laughlin},  {Butler}, {Marcy}, {Vogt}, {Bodenheimer}, {Ida}, {Toyota}, {Wolf}, {Valenti},  {Boyd}, {Johnson}, {Wright}, {Ammons}, {Robinson}, {Strader}, {McCarthy},  {Tah}, \& {Minniti}}]{sato05}
{Sato}, B., {et~al.} 2005, \apj, 633, 465

\bibitem[{{Sato} {et~al.}(2008a){Sato}, {Izumiura}, {Toyota},  {Kambe}, {Ikoma}, {Omiya}, {Masuda}, {Takeda}, {Murata}, {Itoh}, {Ando},  {Yoshida}, {Kokubo}, \& {Ida}}]{sato08a}
{Sato}, B., {et~al.} 2008a, \pasj, 60, 539

\bibitem[{{Sato} {et~al.}(2008b){Sato}, {Toyota}, {Omiya},  {Izumiura}, {Kambe}, {Masuda}, {Takeda}, {Itoh}, {Ando}, {Yoshida}, {Kokubo},  \& {Ida}}]{sato08b}
{Sato}, B., {et~al.} 2008b, \pasj, 60, 1317

\bibitem[{{Schlaufman} \& {Laughlin}(2010)}]{sl10}
{Schlaufman}, K.~C. \& {Laughlin}, G. 2010, \aap, 519, A105+

\bibitem[{{Schlaufman} \& {Laughlin}(2011)}]{sl11}
---. 2011, ArXiv e-prints

\bibitem[{Schwarz(1978)}]{schwarz78}
Schwarz, G. 1978, The Annals of Statistics, 461

\bibitem[{{Sousa} {et~al.}(2008){Sousa}, {Santos}, {Mayor}, {Udry},  {Casagrande}, {Israelian}, {Pepe}, {Queloz}, \& {Monteiro}}]{sousa08}
{Sousa}, S.~G., {et~al.} 2008, \aap, 487, 373

\bibitem[{{Spiegel} {et~al.}(2011){Spiegel}, {Burrows}, \&  {Milsom}}]{spiegel11}
{Spiegel}, D.~S., {Burrows}, A., \& {Milsom}, J.~A. 2011, \apj, 727, 57

\bibitem[{{Takeda} {et~al.}(2008){Takeda}, {Sato}, \& {Murata}}]{takeda08}
{Takeda}, Y., {Sato}, B., \& {Murata}, D. 2008, \pasj, 60, 781

\bibitem[{{Thommes} {et~al.}(2008){Thommes}, {Matsumura}, \&  {Rasio}}]{thommes08}
{Thommes}, E.~W., {Matsumura}, S., \& {Rasio}, F.~A. 2008, Science, 321, 814

\bibitem[{{Thommes} \& {Murray}(2006)}]{thommes06}
{Thommes}, E.~W. \& {Murray}, N. 2006, \apj, 644, 1214

\bibitem[{{Tull}(1998)}]{tull98}
{Tull}, R.~G. 1998, in Society of Photo-Optical Instrumentation Engineers  (SPIE) Conference Series, Vol. 3355, Society of Photo-Optical Instrumentation  Engineers (SPIE) Conference Series, ed. {S.~D'Odorico}, 387--398

\bibitem[{{Tull} {et~al.}(1995){Tull}, {MacQueen}, {Sneden}, \&  {Lambert}}]{tull95}
{Tull}, R.~G., {et~al.} 1995,  \pasp, 107, 251

\bibitem[{{Valenti} {et~al.}(2009){Valenti}, {Fischer}, {Marcy}, {Johnson},  {Henry}, {Wright}, {Howard}, {Giguere}, \& {Isaacson}}]{valenti09}
{Valenti}, J.~A., {et~al.} 2009, \apj, 702, 989

\bibitem[{{Valenti} \& {Fischer}(2005)}]{valenti05}
{Valenti}, J.~A. \& {Fischer}, D.~A. 2005, \apjs, 159, 141

\bibitem[{{Valenti} \& {Piskunov}(1996)}]{valenti96}
{Valenti}, J.~A. \& {Piskunov}, N. 1996, \aaps, 118, 595

\bibitem[{{van Leeuwen}(2007)}]{hipp2}
{van Leeuwen}, F. 2007, \aap, 474, 653

\bibitem[{{VandenBerg} \& {Clem}(2003)}]{vandenberg03}
{VandenBerg}, D.~A. \& {Clem}, J.~L. 2003, \aj, 126, 778

\bibitem[{{Vogt} {et~al.}(1994){Vogt}, {Allen}, {Bigelow}, {Bresee}, {Brown},  {Cantrall}, {Conrad}, {Couture}, {Delaney}, {Epps}, {Hilyard}, {Hilyard},  {Horn}, {Jern}, {Kanto}, {Keane}, {Kibrick}, {Lewis}, {Osborne},  {Pardeilhan}, {Pfister}, {Ricketts}, {Robinson}, {Stover}, {Tucker}, {Ward},  \& {Wei}}]{vogt94}
{Vogt}, S.~S., {et~al.} 1994, in Proc.  SPIE Instrumentation in Astronomy VIII, David L. Crawford; Eric R. Craine;  Eds., Volume 2198, p. 362, ed. D.~L. {Crawford} \& E.~R. {Craine}, 362--+

\bibitem[{{Winn} {et~al.}(2007){Winn}, {Holman}, \& {Fuentes}}]{winn07}
{Winn}, J.~N., {Holman}, M.~J., \& {Fuentes}, C.~I. 2007, \aj, 133, 11

\bibitem[{{Wright}(2005)}]{wright05}
{Wright}, J.~T. 2005, \pasp, 117, 657

\bibitem[{{Wright} \& {Howard}(2009)}]{wrighthoward}
{Wright}, J.~T. \& {Howard}, A.~W. 2009, \apjs, 182, 205

\bibitem[{{Wright} {et~al.}(2004){Wright}, {Marcy}, {Butler}, \&  {Vogt}}]{wright04b}
{Wright}, J.~T., {et~al.} 2004, \apjs,  152, 261

\bibitem[{{Wright} {et~al.}(2009){Wright}, {Upadhyay}, {Marcy}, {Fischer},  {Ford}, \& {Johnson}}]{wright09}
{Wright}, J.~T., {et~al.} 2009, \apj, 693, 1084

\bibitem[{{Wyatt} {et~al.}(2007){Wyatt}, {Clarke}, \& {Greaves}}]{wyatt07}
{Wyatt}, M.~C., {Clarke}, C.~J., \& {Greaves}, J.~S. 2007, \mnras, 380, 1737

\bibitem[{{Yi} {et~al.}(2004){Yi}, {Demarque}, \& {Kim}}]{y2}
{Yi}, S.~K., {Demarque}, P., \& {Kim}, Y.-C. 2004, \apss, 291, 261

\end{thebibliography}

\clearpage
\tabletypesize{\scriptsize}
\begin{deluxetable}{llllllllllllll}
\tabletypesize{\tiny}
\tablecolumns{14}
\tablecaption{Stellar Parameters\label{tab:stars}}
\tablewidth{0pt}
\tablehead{
  \colhead{Star}          &
  \colhead{V}                  &
  \colhead{$B-V$}              &
  \colhead{Distance}      &
  \colhead{$M_V$}              &
  \colhead{${\rm [Fe/H]}$}     &
  \colhead{$T_{\rm eff}$}  &
  \colhead{\vsini}       &
  \colhead{$\log{g}$}          &
  \colhead{$M_{*}$}    &
  \colhead{$R_{*}$}    &
  \colhead{$L_{*}$}    &
  \colhead{Age}          &
  \colhead{$S_{HK}$}           \\
  \colhead{}          &
  \colhead{}                  &
  \colhead{}              &
  \colhead{(pc)}      &
  \colhead{}              &
  \colhead{}     &
  \colhead{(K)}  &
  \colhead{(\ks)}       &
  \colhead{(cgs)}          &
  \colhead{(\msun)}    &
  \colhead{(\rsun)}    &
  \colhead{(\lsun)}    &
  \colhead{(Gyr)}          &
  \colhead{}           \\
  \colhead{(1)}          &
  \colhead{(2)}                  &
  \colhead{(3)}              &
  \colhead{(4)}      &
  \colhead{(5)}              &
  \colhead{(6)}     &
  \colhead{(7)}  &
  \colhead{(8)}       &
  \colhead{(9)}          &
  \colhead{(10)}    &
  \colhead{(11)}    &
  \colhead{(12)}    &
  \colhead{(13)}          &
  \colhead{(14)}           
}
\startdata
\starR                 & 
   \vmagR              &
   \bvR                &
   \dR~(\deR)          & 
   \mvR~(\mveR)        & 
   \feR~(0.03)         &       
   \teffR~(44)         &
   \vsiniR~(0.5)       & 
   \loggR~(0.06)       &
   \mstarR~(\mstareR)      &
   \rstarR~(\rstareR)  &
   \lstarR~(\lstareR)  &
   \ageR~(\ageeR)      &
   \sR                 \\
\starB                 & 
   \vmagB              &
   \bvB                &
   \dB~(\deB)          & 
   \mvB~(\mveB)        & 
   \feB~(0.03)         &       
   \teffB~(44)         &
   \vsiniB~(0.5)       & 
   \loggB~(0.06)       &
   \mstarB~(\mstareB)      &
   \rstarB~(\rstareB)  &
   \lstarB~(\lstareB)  &
   \ageB~(\ageeB)      &
   \sB                 \\
\starC                 & 
   \vmagC              &
   \bvC                &
   \dC~(\deC)          & 
   \mvC~(\mveC)        & 
   \feC~(0.03)         &       
   \teffC~(44)         &
   \vsiniC~(0.5)       & 
   \loggC~(0.06)       &
   \mstarC~(\mstareC)      &
   \rstarC~(\rstareC)  &
   \lstarC~(\lstareC)  &
   \ageC~(\ageeC)      &
   \sC                 \\
\starD                 & 
   \vmagD              &
   \bvD                &
   \dD~(\deD)          & 
   \mvD~(\mveD)        & 
   \feD~(0.03)         &       
   \teffD~(44)         &
   \vsiniD~(0.5)       & 
   \loggD~(0.06)       &
   \mstarD~(\mstareD)      &
   \rstarD~(\rstareD)  &
   \lstarD~(\lstareD)  &
   \ageD~(\ageeD)      &
   \sD                 \\
\starL                 & 
   \vmagL              &
   \bvL                &
   \dL~(\deL)          & 
   \mvL~(\mveL)        & 
   \feL~(0.03)         &       
   \teffL~(44)         &
   \vsiniL~(0.5)       & 
   \loggL~(0.06)       &
   \mstarL~(\mstareL)      &
   \rstarL~(\rstareL)  &
   \lstarL~(\lstareL)  &
   \ageL~(\ageeL)      &
   \sL                 \\
\starE                 & 
   \vmagE              &
   \bvE                &
   \dE~(\deE)          & 
   \mvE~(\mveE)        & 
   \feE~(0.03)         &       
   \teffE~(44)         &
   \vsiniE~(0.5)       & 
   \loggE~(0.06)       &
   \mstarE~(\mstareE)      &
   \rstarE~(\rstareE)  &
   \lstarE~(\lstareE)  &
   \ageE~(\ageeE)      &
   \sE                 \\
\starM                 & 
   \vmagM              &
   \bvM                &
   \dM~(\deM)          & 
   \mvM~(\mveM)        & 
   \feM~(0.03)         &       
   \teffM~(44)         &
   \vsiniM~(0.5)       & 
   \loggM~(0.06)       &
   \mstarM~(\mstareM)      &
   \rstarM~(\rstareM)  &
   \lstarM~(\lstareM)  &
   \ageM~(\ageeM)      &
   \sM                 \\
\starN                 & 
   \vmagN              &
   \bvN                &
   \dN~(\deN)          & 
   \mvN~(\mveN)        & 
   \feN~(0.03)         &       
   \teffN~(44)         &
   \vsiniN~(0.5)       & 
   \loggN~(0.06)       &
   \mstarN~(\mstareN)      &
   \rstarN~(\rstareN)  &
   \lstarN~(\lstareN)  &
   \ageN~(\ageeN)      &
   \sN                 \\
\starO                 & 
   \vmagO              &
   \bvO                &
   \dO~(\deO)          & 
   \mvO~(\mveO)        & 
   \feO~(0.03)         &       
   \teffO~(44)         &
   \vsiniO~(0.5)       & 
   \loggO~(0.06)       &
   \mstarO~(\mstareO)      &
   \rstarO~(\rstareO)  &
   \lstarO~(\lstareO)  &
   \ageO~(\ageeO)      &
   \sO                 \\
\starK                 & 
   \vmagK              &
   \bvK                &
   \dK~(\deK)          & 
   \mvK~(\mveK)        & 
   \feK~(0.03)         &       
   \teffK~(44)         &
   \vsiniK~(0.5)       & 
   \loggK~(0.06)       &
   \mstarK~(\mstareK)      &
   \rstarK~(\rstareK)  &
   \lstarK~(\lstareK)  &
   \ageK~(\ageeK)      &
   \sK                 \\
\starF                 & 
   \vmagF              &
   \bvF                &
   \dF~(\deF)          & 
   \mvF~(\mveF)        & 
   \feF~(0.03)         &       
   \teffF~(44)         &
   \vsiniF~(0.5)       & 
   \loggF~(0.06)       &
   \mstarF~(\mstareF)      &
   \rstarF~(\rstareF)  &
   \lstarF~(\lstareF)  &
   \ageF~(\ageeF)      &
   \sF                 \\
\starP                 & 
   \vmagP              &
   \bvP                &
   \dP~(\deP)          & 
   \mvP~(\mveP)        & 
   \feP~(0.03)         &       
   \teffP~(44)         &
   \vsiniP~(0.5)       & 
   \loggP~(0.06)       &
   \mstarP~(\mstareP)      &
   \rstarP~(\rstareP)  &
   \lstarP~(\lstareP)  &
   \ageP~(\ageeP)      &
   \sP                 \\
\starG                 & 
   \vmagG              &
   \bvG                &
   \dG~(\deG)          & 
   \mvG~(\mveG)        & 
   \feG~(0.03)         &       
   \teffG~(44)         &
   \vsiniG~(0.5)       & 
   \loggG~(0.06)       &
   \mstarG~(\mstareG)      &
   \rstarG~(\rstareG)  &
   \lstarG~(\lstareG)  &
   \ageG~(\ageeG)      &
   \sG                 \\
\starH                 & 
   \vmagH              &
   \bvH                &
   \dH~(\deH)          & 
   \mvH~(\mveH)        & 
   \feH~(0.03)         &       
   \teffH~(44)         &
   \vsiniH~(0.5)       & 
   \loggH~(0.06)       &
   \mstarH~(\mstareH)      &
   \rstarH~(\rstareH)  &
   \lstarH~(\lstareH)  &
   \ageH~(\ageeH)      &
   \sH                 \\
\starI                 & 
   \vmagI              &
   \bvI                &
   \dI~(\deI)          & 
   \mvI~(\mveI)        & 
   \feI~(0.03)         &       
   \teffI~(44)         &
   \vsiniI~(0.5)       & 
   \loggI~(0.06)       &
   \mstarI~(\mstareI)      &
   \rstarI~(\rstareI)  &
   \lstarI~(\lstareI)  &
   \ageI~(\ageeI)      &
   \sI                 \\
\starS                 & 
   \vmagS              &
   \bvS                &
   \dS~(\deS)          & 
   \mvS~(\mveS)        & 
   \feS~(0.03)         &       
   \teffS~(44)         &
   \vsiniS~(0.5)       & 
   \loggS~(0.06)       &
   \mstarS~(\mstareS)      &
   \rstarS~(\rstareS)  &
   \lstarS~(\lstareS)  &
   \ageS~(\ageeS)      &
   \sS                 \\
\starQ                 & 
   \vmagQ              &
   \bvQ                &
   \dQ~(\deQ)          & 
   \mvQ~(\mveQ)        & 
   \feQ~(0.03)         &       
   \teffQ~(44)         &
   \vsiniQ~(0.5)       & 
   \loggQ~(0.06)       &
   \mstarQ~(\mstareQ)      &
   \rstarQ~(\rstareQ)  &
   \lstarQ~(\lstareQ)  &
   \ageQ~(\ageeQ)      &
   \sQ                 \\
\starJ                 & 
   \vmagJ              &
   \bvJ                &
   \dJ~(\deJ)          & 
   \mvJ~(\mveJ)        & 
   \feJ~(0.03)         &       
   \teffJ~(44)         &
   \vsiniJ~(0.5)       & 
   \loggJ~(0.06)       &
   \mstarJ~(\mstareJ)      &
   \rstarJ~(\rstareJ)  &
   \lstarJ~(\lstareJ)  &
   \ageJ~(\ageeJ)      &
   \sJ                 \\
\enddata
\end{deluxetable}

\begin{deluxetable}{ccccccccccc}
\tabletypesize{\footnotesize}
\tablewidth{0pt}
\tablecaption{SUMMARY OF PHOTOMETRIC OBSERVATIONS FROM FAIRBORN OBSERVATORY\label{tab:phot}}
\tablehead{
\colhead{Program} & \colhead{Comparison} & \colhead{Check} & 
\colhead{Date Range} & \colhead{Duration} & \colhead{} & 
\colhead{$\sigma{(V-C)}_V$} & \colhead{$\sigma{(V-C)}_B$} &
\colhead{$\sigma{(K-C)}_V$} & \colhead{$\sigma{(K-C)}_B$} & \colhead{} \\
\colhead{Star} & \colhead{Star} & \colhead{Star} 
& \colhead{(HJD $-$ 2,400,000)} & \colhead{(days)} & \colhead{$N_{obs}$} & 
\colhead{(mag)} & \colhead{(mag)} & \colhead{(mag)} & \colhead{(mag)} &
\colhead{Variability} \\
\colhead{(1)} & \colhead{(2)} & \colhead{(3)} & \colhead{(4)} & \colhead{(5)} &
\colhead{(6)} & \colhead{(7)} & \colhead{(8)} & \colhead{(9)} & \colhead{(10)} &
\colhead{(11)}
}
\startdata
  HD 1502 &   HD 3087 &   HD 3434 & 54756--55578 & 822 & 236 & 0.0044 & 0.0042 & 0.0052 & 0.0037 & Constant \\
  HD 5891 &   HD 5119 &   HD 4568 & 55167--55588 & 421 &  82 & 0.0058 & 0.0045 & 0.0046 & 0.0044 & Constant \\
 HD 18742 &  HD 18166 &  HD 20321 & 55167--55599 & 432 & 216 & 0.0066 & 0.0051 & 0.0076 & 0.0058 & Constant \\
 HD 28678 &  HD 28736 &  HD 28978 & 55241--55637 & 396 & 118 & 0.0037 & 0.0035 & 0.0052 & 0.0035 & Constant \\
 HD 30856 &  HD 30051 &  HD 30238 & 55241--55617 & 376 & 210 & 0.0069 & 0.0057 & 0.0080 & 0.0068 & Constant \\
 HD 33142 &  HD 33093 &  HD 34045 & 55104--55639 & 535 & 341 & 0.0054 & 0.0046 & 0.0063 & 0.0070 & Constant \\
 HD 82886 &  HD 81440 &  HD 81039 & 55128--55673 & 545 & 252 & 0.0045 & 0.0036 & 0.0059 & 0.0060 & Constant \\
 HD 96063 &  HD 94729 &  HD 96855 & 55554--55673 & 119 &  92 & 0.0061 & 0.0037 & 0.0052 & 0.0034 & Constant \\
 HD 98219 &  HD 96483 &  HD 98346 & 55554--55673 & 119 & 162 & 0.0073 & 0.0050 & 0.0085 & 0.0077 & Constant \\
 HD 99706 &  HD 99984 & HD 101620 & 55554--55673 & 119 & 167 & 0.0030 & 0.0032 & 0.0033 & 0.0030 & Constant \\
HD 102329 & HD 101730 & HD 100563 & 55241--55673 & 432 & 169 & 0.0053 & 0.0040 & 0.0046 & 0.0048 & Constant \\
HD 106270 & HD 105343 & HD 105205 & 55241--55671 & 430 & 161 & 0.0059 & 0.0055 & 0.0061 & 0.0059 & Constant \\
HD 108863 & HD 109083 & HD 107168 & 55241--55674 & 433 & 191 & 0.0046 & 0.0037 & 0.0047 & 0.0035 & Constant \\
HD 116029 & HD 116316 & HD 118244 & 55242--55673 & 431 & 181 & 0.0044 & 0.0034 & 0.0048 & 0.0041 & Constant \\
HD 131496 & HD 130556 & HD 129537 & 55242--55673 & 431 & 159 & 0.0044 & 0.0044 & 0.0050 & 0.0046 & Constant \\
HD 152581 & HD 153796 & HD 153376 & 55577--55674 &  97 & 111 & 0.0050 & 0.0053 & 0.0066 & 0.0052 & Constant \\
HD 158038 & HD 157565 & HD 157466 & 55122--55674 & 552 & 155 & 0.0044 & 0.0037 & 0.0044 & 0.0037 & Constant \\
\enddata
\end{deluxetable}

\begin{deluxetable}{llllllllllll}
\tablecolumns{12}
\tablecaption{Orbital Parameters\label{tab:planets}}
\tablewidth{0pt}
\tablehead{\colhead{Planet}               &
  \colhead{Period}                   &
  \colhead{T$_p$\tablenotemark{a}}   &
  \colhead{Eccentricity\tablenotemark{b}}         &
  \colhead{K}                       &
  \colhead{$\omega$}                &
  \colhead{\msini}                & 
  \colhead{$a$}                      &
  \colhead{Linear Trend}  &
  \colhead{rms}                     & 
  \colhead{Jitter}                  &
  \colhead{N$_{\rm obs}$}                 \\
  \colhead{}                  &
  \colhead{(d)}                   &
  \colhead{(HJD-2,440,000)}   &
  \colhead{}                  &
  \colhead{(\ms)}                       &
  \colhead{(deg)}                &
  \colhead{(\mjup)}                & 
  \colhead{(AU)}                      &
  \colhead{(\ms~yr$^{-1}$)}  &
  \colhead{(\ms)}                  &
  \colhead{(\ms)}                 &
  \colhead{}                 \\
  \colhead{(1)}                  &
  \colhead{(2)}                  &
  \colhead{(3)}              &
  \colhead{(4)}      &
  \colhead{(5)}              &
  \colhead{(6)}     &
  \colhead{(7)}  &
  \colhead{(8)}       &
  \colhead{(9)}          &
  \colhead{(10)}    &
  \colhead{(11)}    &
  \colhead{(12)}    
}
\startdata
HD\,1502\,b        &
  \pR~(\peR)          &
  \tpR~(\tpeR)        &
  \eR~(\eeR)          &
  \kR~(\keR)          &
  \omR~(\omeR)        &
  \msiniR~(\msinieR)  &
  \arelR~(\areleR)    &
  0 (fixed)           &
  \rmsR               &
  \jittR~(\jitteR)    &
  \nobsR              \\
HD\,5891\,b           &
  \pB~(\peB)          &
  \tpB~(\tpeB)        &
  \eB~(\eeB)          &
  \kB~(\keB)          &
  \omB~(\omeB)        &
  \msiniB~(\msinieB)  &
  \arelB~(\areleB)    &
  0 (fixed)           &
  \rmsB               &
  \jittB~(\jitteB)    &
  \nobsB              \\
HD\,18742\,b          &
  \pC~(\peC)          &
  \tpC~(\tpeC)        &
   $< $~\eupC &
  \kC~(\keC)          &
  \omC~(\omeC)        &
  \msiniC~(\msinieC)  &
  \arelC~(\areleC)    &
  \trendC~(\trendeC)  &
  \rmsC               &
  \jittC~(\jitteC)    &
  \nobsC              \\
HD\,28678\,b      &
  \pD~(\peD)          &
  \tpD~(\tpeD)        &
  \eD~(\eeD)          &
  \kD~(\keD)          &
  \omD~(\omeD)        &
  \msiniD~(\msinieD)  &
  \arelD~(\areleD)    &
  \trendD~(\trendeD)  &
  \rmsD               &
  \jittD~(\jitteD)    &
  \nobsD              \\
HD\,30856\,b       &
  \pL~(\peL)          &
  \tpL~(\tpeL)        &
   $< $~\eupL &
  \kL~(\keL)          &
  \omL~(\omeL)        &
  \msiniL~(\msinieL)  &
  \arelL~(\areleL)    &
  0 (fixed)           &
  \rmsL               &
  \jittL~(\jitteL)    &
  \nobsL              \\
HD\,33142\,b       &
  \pE~(\peE)          &
  \tpE~(\tpeE)        &
   $< $~\eupE &
  \kE~(\keE)          &
  \omE~(\omeE)        &
  \msiniE~(\msinieE)  &
  \arelE~(\areleE)    &
  0 (fixed)           &
  \rmsE               &
  \jittE~(\jitteE)    &
  \nobsE              \\
HD\,82886\,b       &
  \pM~(\peM)          &
  \tpM~(\tpeM)        &
   $< $~\eupM &
  \kM~(\keM)          &
  \omM~(\omeM)        &
  \msiniM~(\msinieM)  &
  \arelM~(\areleM)    &
  \trendM~(\trendeM)  &
  \rmsM               &
  \jittM~(\jitteM)    &
  \nobsM              \\
HD\,96063\,b       &
  \pN~(\peN)          &
  \tpN~(\tpeN)        &
   $< $~\eupN &
  \kN~(\keN)          &
  \omN~(\omeN)        &
  \msiniN~(\msinieN)  &
  \arelN~(\areleN)    &
  0 (fixed)           &
  \rmsN               &
  \jittN~(\jitteN)    &
  \nobsN              \\
HD\,98219\,b       &
  \pO~(\peO)          &
  \tpO~(\tpeO)        &
   $< $~\eupH &
  \kO~(\keO)          &
  \omO~(\omeO)        &
  \msiniO~(\msinieO)  &
  \arelO~(\areleO)    &
  0 (fixed)           &
  \rmsO               &
  \jittO~(\jitteO)    &
  \nobsO              \\
HD\,99706\,b       &
  \pK~(\peK)          &
  \tpK~(\tpeK)        &
  \eK~(\eeK)          &
  \kK~(\keK)          &
  \omK~(\omeK)        &
  \msiniK~(\msinieK)  &
  \arelK~(\areleK)    &
  \trendK~(\trendeK)  &
  \rmsK               &
  \jittK~(\jitteK)    &
  \nobsK              \\
HD\,102329\,b      &
  \pF~(\peF)          &
  \tpF~(\tpeF)        &
  \eF~(\eeF)          &
  \kF~(\keF)          &
  \omF~(\omeF)        &
  \msiniF~(\msinieF)  &
  \arelF~(\areleF)    &
  0 (fixed)           &
  \rmsF               &
  \jittF~(\jitteF)    &
  \nobsF              \\
HD\,106270\,b\tablenotemark{c}      &
  \pP~(\peP)          &
  \tpP~(\tpeP)        &
  \eP~(\eeP)          &
  \kP~(\keP)          &
  \omP~(\omeP)        &
  \msiniP~(\msinieP)  &
  \arelP~(\areleP)    &
  0 (fixed)           &
  \rmsP               &
  \jittP~(\jitteP)    &
  \nobsP              \\
HD\,108863\,b      &
  \pG~(\peG)          &
  \tpG~(\tpeG)        &
   $< $~\eupG &
  \kG~(\keG)          &
  \omG~(\omeG)        &
  \msiniG~(\msinieG)  &
  \arelG~(\areleG)    &
  0 (fixed)           &
  \rmsG               &
  \jittG~(\jitteG)    &
  \nobsG              \\
HD\,116029\,b      &
  \pH~(\peH)          &
  \tpH~(\tpeH)        &
   $< $~\eupH &
  \kH~(\keH)          &
  \omH~(\omeH)        &
  \msiniH~(\msinieH)  &
  \arelH~(\areleH)    &
  0 (fixed)           &
  \rmsH               &
  \jittH~(\jitteH)    &
  \nobsH              \\
HD\,131496\,b      &
  \pI~(\peI)          &
  \tpI~(\tpeI)        &
  \eI~(\eeI)          &
  \kI~(\keI)          &
  \omI~(\omeI)        &
  \msiniI~(\msinieI)  &
  \arelI~(\areleI)    &
  0 (fixed)           &
  \rmsI               &
  \jittI~(\jitteI)    &
  \nobsI              \\
HD\,142245\,b       &
  \pS~(\peS)          &
  \tpS~(\tpeS)        &
   $< $~\eupS &
  \kS~(\keS)          &
  \omS~(\omeS)        &
  \msiniS~(\msinieS)  &
  \arelS~(\areleS)    &
  0 (fixed)           &
  \rmsS               &
  \jittS~(\jitteS)    &
  \nobsS              \\
HD\,152581\,b      &
  \pQ~(\peQ)          &
  \tpQ~(\tpeQ)        &
   $< $~\eupQ &
  \kQ~(\keQ)          &
  \omQ~(\omeQ)        &
  \msiniQ~(\msinieQ)  &
  \arelQ~(\areleQ)    &
  0 (fixed)           &
  \rmsQ               &
  \jittQ~(\jitteQ)    &
  \nobsQ              \\
HD\,158038\,b      &
  \pJ~(\peJ)          &
  \tpJ~(\tpeJ)        &
  \eJ~(\eeJ)          &
  \kJ~(\keJ)          &
  \omJ~(\omeJ)        &
  \msiniJ~(\msinieJ)  &
  \arelJ~(\areleJ)    &
  \trendJ~(\trendeJ)  &
  \rmsJ               &
  \jittJ~(\jitteJ)    &
  \nobsJ              \\
\enddata
\tablenotetext{a}{Time of periastron passage.}
\tablenotetext{b}{When the measured eccentricity is consistent with
  $e=0$ within $2\sigma$, we quote the 2-$\sigma$ upper limit from the
  MCMC analysis.} 
\tablenotetext{c}{One possible orbit solution is reported here, along
  with the formal uncertainties reported by the MCMC analysis. See
  \S~\ref{results} for a note on this system.}
\end{deluxetable}

\end{document}